\documentclass[aps,pre,twocolumn,superscriptaddress,showpacs,showkeys,nofootinbib]{revtex4-2}

\usepackage[svgnames]{xcolor}
\usepackage{graphicx}                   
\usepackage{amsmath,amssymb}            
\usepackage{comment}
\usepackage{subfigure}					
\usepackage[colorlinks=true,
            linkcolor=blue,
            citecolor=blue,
            urlcolor=blue]{hyperref}

\begin{document}
	   
\title{Emergent Topological Complexity in the Barabási–Albert Model with Higher-Order Interactions}

\author{V. Adami}
\affiliation{Department of Physics, University of Mohaghegh Ardabili, P.O. Box 179, Ardabil, Iran}
\author{H. Masoomy}
\affiliation{Department of Physics, Shahid Beheshti University,  1983969411, Tehran, Iran}

\author{M. Lukovi\'c}
\affiliation{Swiss Federal Institute for Forest, Snow and Landscape Research, WSL, 8903 Birmensdorf, Switzerland }

\author{M. N. Najafi}
\email{morteza.nattagh@gmail.com}
\affiliation{Department of Physics, University of Mohaghegh Ardabili, P.O. Box 179, Ardabil, Iran}

\begin{abstract}
We examine the homological structure of the Barabási–Albert model, focusing on the time evolution of $\Delta$-dimensional simplices and topological holes as functions of time $t$ and the attachment parameter $m$ (the number of edges added by each incoming node). Numerical simulations reveal a non-trivial topological transition (TT) in the $(\Delta, m)$ plane, marking a change from a topologically trivial regime to non-trivial topology. This transition signals the emergence of topological complexity in the model, where higher-order structures develop self-similarly across scales. Beyond this transition, the network exhibits self-similar topological growth, evidenced by a power-law decay in the increments of $\Delta$-simplices with $m$-dependent exponents. An analogous transition occurs in the Betti numbers, which display self-similar behavior near the TT and an arctangent dependence farther from it. Based on simulation data, we propose explicit scaling relations describing the behavior of both $\Delta$-simplices and Betti numbers near the TT. Overall, the analysis reveals a rich, gapful topological transition structure, where topological quantities exhibit discrete jumps at the transition point.
\end{abstract}  


\maketitle
\tableofcontents

\section{Introduction}
Complex networks have emerged as a fundamental framework for understanding the structure and dynamics of a wide range of systems, from social and biological interactions to technological and information networks~\cite{albert2002statistical,BOCCALETTI2006175,NewmanS003614450342480}. Among various generative models, the Barabási–Albert (BA) model~\cite{barabasi1999emergence} occupies a central role due to its ability to reproduce the scale-free networks observed in many real-world systems. While the BA model has been extensively studied in terms of its degree correlations, clustering properties, and growth dynamics~\cite{pastor2001dynamical,fotouhi2013degree,szabo2003structural,krapivsky2000connectivity}, recent advances in topological data analysis and algebraic topology have opened new avenues for characterizing complex networks through their higher-order structures beyond pairwise connectivity~\cite{horak2009persistent,sizemore2018cliques,millan2020explosive,iacopini2019simplicial,giusti2016two}. These approaches extend traditional graph theory by incorporating simplicial complexes, which encode multi-node interactions as higher-dimensional simplexes, thus enabling a richer characterization of network topology. The study of simplicial complexes associated with networks has revealed that higher-order structures play a crucial role in phenomena such as synchronization~\cite{millan2020explosive}, contagion dynamics~\cite{iacopini2019simplicial}, and percolation~\cite{zhao2022higher}. 

Topological descriptors such as Betti numbers quantify the number of $\Delta$-dimensional holes, providing a natural way to track the emergence and disappearance of connected components, loops, and voids as a function of a filtration parameter~\cite{horak2009persistent,edelsbrunner2002topological,edelsbrunner2010computational}. Topological properties are important not only because they can characterise the underlying structure of a system but because they can also tell us something about how the structure determines its function. This is especially true when we consider neural networks in which topological voids play an indispensable role in brain functioning \cite{shi2021}.  According to the authors of Ref.~\cite{reimann2017}, cliques describe the flow of information in neural networks at the local level, while cavities provide a global measure of information flow in the whole network. Cavities are distributed almost everywhere in the brain, connecting many different regions together \cite{shi2021}. More discussions about the importance of topology in neuroscience can be found in Ref.~\cite{giusti2016}. On the other hand, lower order topological voids have been used in simpler systems to characterise protein compressibility \cite{gameiro2015}, the heterogeneity of rock pore geometry in geological formations \cite{jiang2018}, similarity of pore structures in general \cite{yongjin2017} and bonding trends encoded in various electron densities in crystals \cite{szymanski2025}. In principle, complex systems such as the brain network have a much richer inherent structure due to the fact that multi-interactions among simplicial complexes are allowed \cite{serrano2020centrality}.

In growing network models, time can be treated as a natural filtration variable, allowing the evolution of topological invariants to be followed dynamically~\cite{horak2009persistent,Siu2025,petri2014homological}. This perspective links network growth processes to the evolution of topological features and enables the identification of scaling laws and critical phenomena analogous to phase transitions in statistical physics~\cite{horak2009persistent,courtney2016generalized,sizemore2017classification}. Despite growing interest in network topology, the temporal evolution of simplicial complexes in the BA model remains relatively unexplored. Most prior studies have focused on static properties of network topology or on persistent homology derived from fixed configurations~\cite{horak2009persistent,Siu2025,Adami_2025}. However, understanding how higher-dimensional simplexes and holes evolve with time provides deeper insight into the generative mechanisms that govern complex network formation. Specifically, the interplay between preferential attachment and topological growth may lead to emergent scaling behavior~\cite{courtney2016generalized,bianconi2016network,bianconi2017emergent,kovalenko2021growing}, critical thresholds~\cite{KAHLE20091658,sizemore2019importance}, and phase-like transitions across different dimensions of the simplicial complex~\cite{Siu2025,kahle2014sharp}.

In this work, we investigate the dynamical behavior of $\Delta$-dimensional simplexes and holes in the simplicial complexes constructed from the BA model, with the network growth time $t$ serving as the filtration parameter. Through extensive simulations, we analyze the evolution of Betti numbers up to dimension six and study their dependence on time and the connectivity parameter $m$, which controls the number of new connections added per node. Our results reveal that the number of $\Delta$-dimensional simplexes exhibits robust scaling laws with time, indicative of self-similar growth patterns in the higher-order topology of the BA network. Interestingly, the dynamics of $\Delta$-dimensional holes show a distinct nontrivial temporal dependence that can be fitted good by an arctangent function.

Furthermore, we identify signatures of topological transitions in the ($m$-$\Delta$) parameter space. These transitions manifest as abrupt changes both in the simplexes and Betti numbers, marking the onset of qualitatively different topological regimes. Such behavior highlights the complex interplay between network connectivity and higher-order structure formation.

\section{Structure and Achievements}
In this section, we explore the structure and the main findings of this study to facilitate reading the paper. The next section is devoted to topological concepts in complex networks, where we explore the core definitions essential for the topology of growing networks. Section~\ref{sec:simplicies} present some mean-field arguments and simulation results for the system.

Our findings are listed as follows:
\begin{itemize}
    \item In the mean-field level, the probability of formation of a $\Delta$ simplex is given by Eqs.~\ref{Eq:SigmaMF1} and~\ref{Eq:SigmaMF2}. We argue that the deviations from mean-field results are considerable for large $\Delta$ values.
    \item Our simulations (Figs.~\ref{fig:sigma_F_psi} and~\ref{fig:sigma_F_psi_2}) reveal that the increment of the simplicial complexes vary with time according to Eq.~\ref{eqt:sigma_delta_t_m}, with a proportionality parameter given in Eq.~\ref{eq:F_m}. The fitting parameters are reported in table~\ref{tab:fitting_params}.
    \item A phase diagram is sketched in Fig.~\ref{fig:m_delta_simplex} which shows the two dimensional ($\Delta,m$) dependence of the cumulated number of simplicial complexes. It shows explicitly how the topological phases are separated by a straight transition line.
    \item The Betti numbers are conjectured to behave according to Eq.~\ref{Eq:betti_vs_t} with respect to time. These solutions follow an arctan function, and therefore saturate asymptotically to a finite value. The proportionality parameter shows a power-law behavior according to Eq.~\ref{eq_betti_infinity_fitting}, the fitting parameters of which are shown in Fig.~\ref{fig:betti_parameters}. The corresponding topological phase space is sketched in Fig.~\ref{fig:total_number_of_hole}, where the blank circles refer to trivial topologies, while the colored ones refer to non-trivial topologies with higher dimensional voids in the network.
\end{itemize}
The rest of the paper deals with expanding and analyzing the proposed relations. We conclude the paper with some remarks in section~\ref{SEC:conclusion}.

\section{Complex Networks and Topological Concepts}\label{SEC:tda_simple}

Traditional complex network analysis relies on pairwise interactions, representing systems as graphs whose edges encode relationships between two entities at a time. While this abstraction has been enormously successful, it inevitably compresses richer multiway interactions into binary links, obscuring essential structural and dynamical information. Many real-world systems, including social groups, biochemical complexes, co-authorship teams, neuronal assemblies~\cite{albert2002statistical,BOCCALETTI2006175,NewmanS003614450342480}, operate through simultaneous interactions among three or more units, not reducible to independent dyads. Network measures based solely on edges (degree, clustering, assortativity) miss higher-order connectivity patterns and can misrepresent the system’s true interaction geometry.

Higher-order interaction analysis provides a new insight to these systems by extending pairwise interactions to multiple interactions in simplicial complexes, where nodes, edges, triangles, tetrahedra, and higher-dimensional simplexes are the essential objects to be analyzed. Among the other tools, topology as a mathematical tool for studying global properties of objects especially under the continuous transformations, offers a rigorous framework for probing structural features that persist beyond local, pairwise interactions. Rather than emphasizing particular link configurations, topology focuses on properties that remain invariant under smooth deformations~\cite{nakahara2018geometry}. Topological data analysis (TDA) offers the mathematical machinery—such as persistent homology and Betti numbers—to quantify the formation of higher-dimensional holes, cavities, and voids across scales or growth processes~\cite{nakahara2018geometry}.

In this setting, topological invariants—including the number of connected components (0-dimensional features), cycles or loops (1-dimensional features), and higher-dimensional cavities or voids (2D, 3D, and beyond)—provide a principled way to classify and compare complex systems. These quantities encode the homological structure of networks, revealing aspects of robustness, hierarchy, and multi-scale organization that cannot be inferred from edge-based measures alone. 

\subsection{Simplicial Complexes; Homology and Betti Numbers for Networks}
Although networks are not geometric objects in the classical sense, they can be endowed with a topological structure through the construction of simplicial complexes, which systematically incorporate higher-order interactions by assembling nodes, edges, triangles, and higher-dimensional simplexes into a coherent combinatorial space.\\

A \textit{simplicial complex} $K$ is a collection of simplexes (nodes, edges, triangles, tetrahedra, etc.) that satisfies two key properties: (I) If a simplex $\Sigma \in K$, then every face of $\Sigma$ is also in $K$. (II) The intersection of any two simplexes in $K$ is either empty or a face common to both. These rules ensure that simplexes assemble into a coherent combinatorial structure that consistently encodes higher-order relationships.

In this setting, nodes form 0-simplexes, edges form 1-simplexes, and multiway interactions such as triangles or tetrahedra appear as 2- and 3-simplexes, with the construction extending naturally to higher dimensions. This representation captures genuine group interactions rather than reducing them to pairs, enabling the study of higher-order connectivity patterns that graphs cannot express. Simplicial complexes also serve as the foundation for computing topological invariants---such as Betti numbers---that describe connected components, loops, and higher-dimensional cavities. 

Given a simplicial complex, one can compute \textit{homology}, which captures different types of ``holes'' in the network: $\beta_0$ measures number of connected components. $\beta_1$ counts number of independent loops or cycles. $\beta_2$ determines number of voids, etc. These are called the \textit{Betti numbers}, and they summarize the topological structure of the network at different scales.

\subsection{Filtration and Persistent Homology} \label{SEC:persistent_homology_simple}

To study how a network's topology evolves we use a process called \textit{filtration}~\cite{edelsbrunner2002topological}. This involves constructing a sequence of simplicial complexes based on a filtration parameter \( w \), such as connection strength or distance:
\begin{equation}
    K(w) \subset K(w + \Delta w).
\end{equation}
As \( w \) increases, new simplices (edges, triangles, etc.) are ``born", and topological features may appear or disappear. This is where \textit{persistent homology} comes in: it tracks how long features \textit{persist} throughout the filtration. Each topological feature is assigned a \textit{persistence pair}:$(w_{\text{birth}}, w_{\text{death}})$ indicating when the feature appears and disappears. These pairs can be visualized using: 

\begin{itemize}
	\item \textit{Persistence diagrams}: points on a 2D persistence space.
	\item \textit{Persistence barcodes}: horizontal bars showing feature lifetimes.
\end{itemize}

In our study, we adopt a dynamic filtration approach by using the time step $t$ as the filtration parameter. This is particularly suitable for growing real networks and generative models, which evolve through specific attachment mechanisms~\cite{barabasi1999emergence,albert2002statistical,krapivsky2000connectivity,Dorogovtsev01062002}.  During filtration, certain topological voids may form or vanish. Characteristics that remain over time tend to reflect genuine aspects of the space, rather than artifacts of our sampling, measurement, or parameter selection methods~\cite{carlsson2009topology}. These models inherently produces a sequence of network states indexed by time \( t \), with each state corresponding to a snapshot of the network at a given size. A dynamical network is defined by a time-dependent graph $G(t)=G(V(t),E(t))$ at each time step $t \in \{1,...,N\}$, where $V(t)$ symbolizes the set of nodes that grow with time, during each of which a new link is added to the set of edges $E(t)$, which connects to existing nodes. By interpreting this time evolution as a filtration:
\begin{equation}
    G(t) \subset G(t + 1) \subset \cdots \subset G(T)
\end{equation}
we construct a corresponding sequence of simplicial complexes:
\begin{equation}
    K(t) \subset K(t + 1) \subset \cdots \subset K(T)
    \label{eqt_filtration}
\end{equation}
where \( K(t) \) represents the simplicial complex formed from the network at time step \( t \). This approach allows us to track the birth and death of topological features—such as connected components and cycles—as the network grows.\\

In the second part of the paper, we focus on $\Delta$-dimensional ($\Delta\geq1$) topological holes, which are studied in the context of homology groups of networks. For example a 1-hole (or loop) is formed by a group of nodes (more than three), each of which is connected to a neighboring node without a link breaking up the loop in two pieces. Apart from the interest in the structure of the network, the appearance of such topological objects affects directly the dynamics that is defined on top of the network. Both the geometrical as well as the topological properties of these networks are important in characterizing the original system being modeled.

Betti numbers are fundamental invariants in algebraic topology that provide insights into the structure of a topological space. They are particularly useful for studying ``holes" of various dimensions. Betti numbers are non-negative integers that count the number of independent topological features (holes) of a space in different dimensions. They are associated with the ranks of homology groups of the space. The $\Delta$-th Betti number is typically denoted as $\beta_{\Delta}$ such that $\beta_0$, $\beta_1$, and $\beta_2$ respectively for instance, count the number of connected components, the number of independent loops or circular holes, and the number of enclosed voids or cavities of the space. 

\subsection{Topology of Scale-Free Networks}

A large class of natural and man-made networks are scale-free in the sense that some scaling behavior (like power-law distributions) is found to govern the quantities of interest, with scaling exponents that classify the networks and define the \textit{universality}. It is normally defined as the networks for which the distribution function of the degree of nodes ($k$) behaves like~\cite{barabasi1999emergence,adami2026centrality}
\begin{equation}
    p(k)\propto k^{-\gamma},
    \label{Eq:PLD}
\end{equation}
where $\gamma$ is the degree exponent. The Internet, the World Wide Web, citation networks, and some social networks are examples of scale-free networks, which include some nodes with unusually high centralities (degree) compared to the rest of the nodes~\cite{albert2002statistical,BOCCALETTI2006175,NewmanS003614450342480,adami2024centrality,vazquez2001large}. \\

The studies~\cite{kovalenko2021growing, qi2025iterative, xie2021higher} demonstrate that incorporating higher-order structures into scale-free networks provide a more complete understanding of both their architecture and dynamics. From a structural perspective, generative models based on simplicial complexes extend preferential attachment to include higher-order interactions, producing networks with scale-free degree distributions and, depending on the growth rules, scale-free generalized degree distributions~\cite{kovalenko2021growing}. Iterative constructions of higher-order fractal networks further show that scale-free behavior can coexist with fractal and hierarchical organization when networks are modeled as pure simplicial complexes~\cite{qi2025iterative}. Beyond topology, higher-order interactions significantly affect dynamical processes on scale-free networks, as illustrated by evolutionary game studies in which second-order neighborhood structures and higher-order preference mechanisms alter cooperative behavior in a manner dependent on the temptation to defect~\cite{xie2021higher}. Overall, these results highlight that higher-order organization is essential for capturing both the structural complexity and dynamical properties of scale-free networks.\\

Although topological properties of growing networks have been extensively investigated (see~\cite{kannan2019persistent, roy2020forman, horak2009persistent, kovalenko2021growing}), the current literature still lacks a comprehensive and systematic study that also includes algebraic topology.

\section{The Barabási–Albert Model}\label{sec:simplicies}

The Barabasi-Albert (BA) model provides a popular mathematical framework for generating complex scale-free networks using a preferential attachment mechanism (Ref~\cite{albert2002statistical} for review) with a power-law behavior according to Eq.~\ref{Eq:PLD}. This model uses a dynamical preferential attachment method, according to which a new node is more likely to attach to nodes with higher degree values. The graph of the dynamical network $G_{\text{BA}}(t,m)$ grows by one node $v_{t} \in V(t)$ at each time step with $m$ new links $e_{t} \equiv \{e_{t,t'} ~ | ~ A_{t,t'}=1\}_{t'=1}^{t-1} \subseteq E(t)$, $|e_{t}|=m$, where $e_{t,t'}$ denotes the edge between the node $v_t$ and the node $v_{t'}$ and $A_{t,t'}=A_{t',t}$ is the corresponding adjacency matrix up to time $t>t'$. Note that the nodes are labeled uniquely by the time of creation. This new random connection to existing nodes is according to their degree centralities: the probability of connection for the node $v_t$ to an already existing node $v_{t'}$, depends on the degree of $v_{t'}$ at time $t>t'$, denoted by $k_{t'}(t)$, via the formulae
\begin{equation}
	p(A_{t,t'}=1) = \frac{k_{t'}(t)}{\sum_{\tau=1}^{t-1} k_{\tau}(t)}.
\end{equation}
The $m$-BA network grown up to the maximum time $t_{\text{max}}\equiv N$ is denoted by $G_{\text{BA}}(N,m)\equiv G_{\text{BA}}(t_{\text{max}},m)$. The $m$-BA model has intensely been investigated from many aspects~\cite{gomez2004local, lee2006statistical, holme2002growing}. Mean-field analysis of the BA model shows that the degree exponent of $m$-BA model is $\gamma_{\text{BA}}=3$ for all values of $m$, which has also been shown numerically to be consistent~\cite{barabasi1999mean}. \\

It is notable that the probability for the existence of a link between node $v_{t_i}$ and node $v_{t_j}$ in a BA model is given by the expression~\cite{klemm2002growing}
\begin{equation}
    P(v_{t_i},v_{t_j})=\frac{m}{2\sqrt{t_it_j}},
    \label{Eq:prob_ij}
\end{equation}
where $t_i$ and $t_j$ are two time steps. This decrease of the probability of forming a link between nodes with time reflects the fact that, during the initial stages of network growth, the pool of existing nodes available for attachment is comparatively small. \\

\textbf{Numerical details: } In this study, we have useed NetworkX package to generate $m$-BA samples. The networks were evolved up to a final network size of $N=10^4$, using connection parameters in the range $1 \leq m \leq 29$. Each observable was determined by averaging over an ensemble of $10^3$ networks.

\subsection{Simplicial Complexes of BA model}
In this paper the higher-order structures of the BA network are studied from the aspect of simplicial complexes and homological properties. To this end, we map the simulated BA networks to growing simplicial complexes $K_{\text{BA}}(t,m)\equiv K(G_{\text{BA}}(t,m))$ with time as the filtration parameter, and analyse the evolution of their simplicial and homological building blocks. To avoid unnecessary complications in notation, we remove the subscript BA, and use the notation $K(t,m)$ for the simplicial complexes and other functions in the remainder of the paper.\\

We analyse the properties of $\Delta$-simplexes of the evolving BA network as a function of $m$, $t$, and the dimension of the simplexes $\Delta$. Due to the absence of rewiring, the BA network has the local temporal topological structure property, according to which adding a new node does not alter the number and structure of the pre-existing $\Delta$-simplexes (not the homological properties) from earlier times. In light of this property, we start by analysing the increments in the number of $\Delta$-simplexes, denoted by $\delta\Sigma_\Delta(t,m)$, that are created between two consecutive time steps $t-1$ and $t$ (see Eq.~\eqref{eqt_filtration}):
\begin{equation*}
    \delta \Sigma_{\Delta}(t,m) = \Sigma_{\Delta}(t,m) - \Sigma_{\Delta}(t-1, m),
    \label{eqt:delta_sigma}
\end{equation*}
where $\Sigma_{\Delta}(t,m)\equiv \Sigma_{\Delta}\left(K(t, m)\right)$ is the total number of $\Delta$-dimensional simplexes present in the network at time step $t$. Note that the total number of simplexes of dimension $\Delta$ created after $N$ time steps, is given by the sum
\begin{equation}
	\Sigma_{\Delta}(N,m) \equiv \sum_{t=1}^{N}\delta\Sigma_{\Delta}(t,m).
	\label{eq:total_simplexes}
\end{equation}
This equation represents the total number of $\Delta$-dimensional simplexes at the end of the network growth process, after time-step $t=t_{max}=N$.  The dependency of a $\Delta$-dimensional simplex on the number of connections $m$ is trivial, since an increase in $m$ leads to an increase in the number of $\Delta$-simplexes for all $\Delta\geq0$. For every time step $t$, after adding a node to the network, we calculated the number of newly created simplexes. 

\subsection{Mean-Field Theory for the Simplicial Complexes}

\begin{figure*}
    \includegraphics[width = 0.95\textwidth]{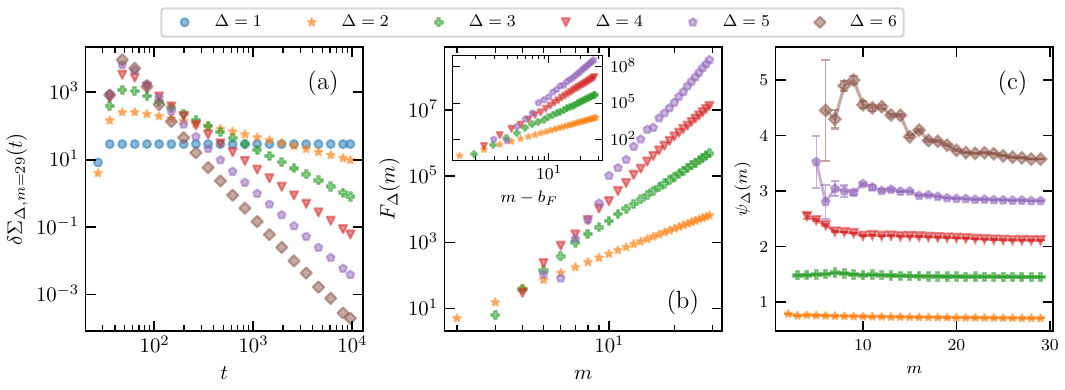}
    \caption{(a) The increments in the number of $\Delta$-dimensional simplexes, $\delta\Sigma_{\Delta,m}$, display a power-law scaling with time $t$ for various dimensions at $m = 29$. (b) and (c) present the corresponding intercepts ($F$) and slopes ($\psi$) of this scaling behavior as functions of $m$ and $\Delta$. For a fixed $\Delta$, the values of $\psi$ approach constant asymptotic limits as $m$ increases.}
    \label{fig:sigma_F_psi}
\end{figure*}

\begin{figure*}
    \includegraphics[width = 0.95\textwidth]{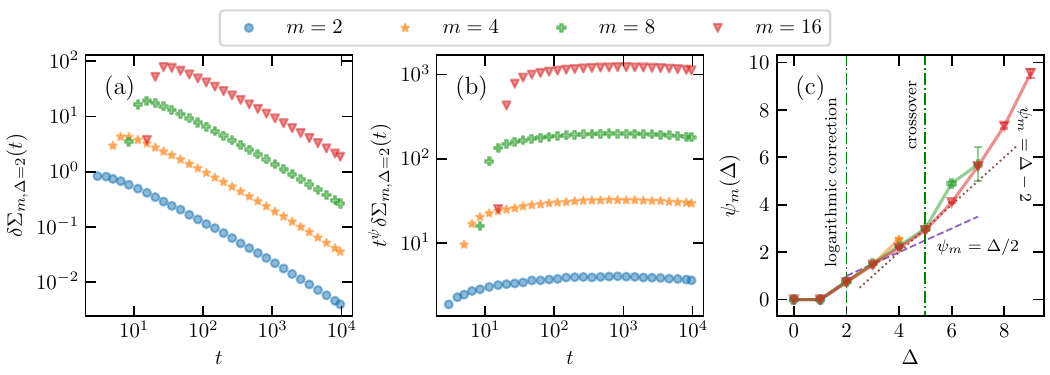}
    \caption{(a) The increments in the number of $\Delta$-dimensional simplexes, $\delta\Sigma_{\Delta,m}$, exhibit a power-law scaling with time $t$ for various dimensions at $\Delta = 2$. (b) To verify the time-independence of the function $F$ defined in Eq.~\eqref{eqt:sigma_delta_t_m}, both sides of the equation are multiplied by $t^{\psi}$, confirming that $F$ remains constant over time. (c) The dependence of the scaling exponent $\psi$ on $(m, \Delta)$ is illustrated. For clarity, the plots are shown for only four representative values of $m$.}
    \label{fig:sigma_F_psi_2}
\end{figure*}

Before presenting the simulation results, we introduce a mean-field (MF) theory. Let us consider a $\Delta$-simplex, which contains $\Delta+1$ nodes and $\frac{1}{2}\Delta(\Delta+1)$ edges. Given the probability of connection in Eq.~\eqref{Eq:prob_ij}, one can express the probability of formation of a $\Delta$-simplex as
\begin{equation}
P_\Delta(t,m)\propto \sum_{\left\{ t_{i_k} \right\}} \prod_{\left\{ m,n \right\}} P\!\left(v_{t_{i_m}} v_{t_{i_n}}\right),
\end{equation}
where the summation runs over all times up to $t$ for the $k=1,2,\ldots,\Delta+1$ nodes, with the restriction that they are all distinct (i.e., the nodes cannot coincide). The product inside the summation is taken over all edges $(m,n)$ belonging to the simplex. 

This MF theory is based on the probability for the existence of a link between two nodes, as given in Eq.~\eqref{Eq:prob_ij}. It implies that the probability of the existence of a $\Delta$-simplex, $P_\Delta$, is given by the product (see Appendix~\ref{SEC:Wicks})

\begin{widetext}
\begin{equation}
\begin{split}
P_\Delta(t,m)&\propto\left(\frac{m}{2}\right)^{\frac{1}{2}\Delta(\Delta+1)}\sum_{t_{i_1}=1}^t\sum_{t_{i_2}\ne t_{i_1}}^t\sum_{t_{i_3}\ne t_{i_2}\ne t_{i_2}}^t...\sum_{t_{i_{\Delta+1}}\ne t_{i_1}...\ne t_{\Delta}}^t\frac{1}{t_{i_1}^{\frac{\Delta}{2}}}\frac{1}{t_{i_2}^{\frac{\Delta}{2}}}\frac{1}{t_{i_3}^{\frac{\Delta}{2}}}...\frac{1}{t_{i_{\Delta+1}}^{\frac{\Delta}{2}}}\\
&=\left(\frac{m}{2}\right)^{\frac{1}{2}\Delta(\Delta+1)}\sum_{t_{i_1},t_{i_2},t_{i_3},...,t_{i_{\Delta+1}}=1}^t\left(t_{i_1}t_{i_2}t_{i_3}...t_{i_{\Delta+1}}\right)^{-\frac{\Delta}{2}}\prod_{i, j=1}^{\Delta+1}\left(1-\delta_{t_i,t_j}\right).
\end{split}
\label{Eq:ProbSigma}
\end{equation}
\end{widetext}
This proportionality relation allows us to ignore the prefactor $\left(\frac{m}{2}\right)^{\frac{1}{2}\Delta(\Delta+1)}$, since it can be absorbed into the normalization constant. In Appendix~\ref{SEC:Wicks}, we develop a Wick-like expansion for this quantity, which enables us to expand the expression and solve it exactly in terms of the harmonic number of order~$r$:
\begin{equation}
	H(t,r)=\sum_{x=1}^t x^{-r}.
	\label{Eq:HarmonicNumber}
\end{equation}
Some examples are 
\begin{equation}
	\begin{split}
		&P_1(t,m)\propto H\left(t,\frac{1}{2}\right)^2-H(t,1),\\
		&P_2(t,m)\propto H(t,1)^3-3H(t,2)H(t,1)+2H\left(t,\frac{3}{2}\right),\\
		&P_3(t,m)=H\left(t,\frac{3}{2}\right)^4-6H(t,3)H\left(t,\frac{3}{2}\right)^2\\
		&+8H\left(t,\frac{9}{2}\right)H\left(t,\frac{3}{2}\right)+3H(t,3)^2-26H(t,6),
	\end{split}
	\label{Eq:sigma1,2,3}
\end{equation}
with which, we can find the asymptotic behavior of $\delta\Sigma_{\Delta}(t)\equiv \partial_tP_{\Delta}(t)$ and also $\Sigma_{\Delta}(t)\propto P_\Delta(t)$ for long times. This analysis is carried out exactly in Appendix~\ref{SEC:Wicks}, where a complete treatment is developed. The leading contribution to the asymptotic scaling behavior is given by
\begin{equation}
	\left. P_{\Delta}(t,m)\right|_{\text{leading term}}
	\propto H\!\left(t,\frac{\Delta}{2}\right)^{\Delta+1},
	\label{Eq:approximationMain}
\end{equation}
which corresponds to neglecting the restriction in the summations, i.e.,
$\prod_{i,j=1}^{\Delta+1}\left(1-\delta_{t_i,t_j}\right)\to 1$,
as a leading-order approximation. This approximation is valid for not too large values of $\Delta$.
The asymptotic scaling behavior of $\delta\Sigma_{\Delta}(t)$ is derived in Appendix~\ref{SEC:Wicks} and reads
\begin{equation}
	\delta\Sigma_\Delta(t\to\infty,m)\propto
	\begin{cases}
		t^{-\psi_{\text{MF}1}(\Delta)}, & \Delta\neq 2,\\[4pt]
		t^{-1}(\ln t)^2, & \Delta=2,
	\end{cases}
	\label{Eq:SigmaMF1}
\end{equation}
where
\begin{equation}
	\psi_{\text{MF}1}(\Delta)=
	\begin{cases}
		0, & \Delta=0,1,\\[4pt]
		\dfrac{\Delta}{2}, & \Delta\ge 3.
	\end{cases}
	\label{Eq:Exponents1}
\end{equation}

We find that, at the mean-field level, the power-law exponents with respect to time are independent of $m$, a result that is confirmed later. While this estimate of the exponents is very accurate for small values of $\Delta$, it deviates from the simulation results for large $\Delta$ ($\Delta \gtrsim 6$). This discrepancy indicates that the approximation in Eq.~\eqref{Eq:approximationMain} becomes unreliable for large $\Delta$, due to strong fluctuations in the establishment of links.

To address this issue, we introduce a second mean-field scheme (MF2), also referred to as the \textit{combinatorial MF theory}. This approach is appropriate for regimes in which attachment fluctuations are large, or equivalently, attachments are more uniform. We formulate a master equation governing the attachment probabilities, under the assumption that the probability of attachment to any node is determined by the average degree of the network. This framework contrasts with the previous MF theory, which assumes small fluctuations. Further details are provided in Appendix~\ref{SEC:Combinatorial}.

To form a $\Delta$-simplex when a new node $v_t$ is added to the system at time step $t$, the new node (with its $m$ links) must attach to all nodes of an existing $(\Delta-1)$-simplex. In this case, $\delta\Sigma_\Delta(t,m)=1$. If the $m$ links of the new node attach to two distinct $(\Delta-1)$-simplices, then $\delta\Sigma_\Delta(t,m)=2$, and so on. Such a connection must be all-to-all, meaning that it must include at least $\Delta$ links to the $(\Delta-1)$-simplex. Consequently, there is a combinatorial factor $\binom{m}{\Delta}$ representing the number of distinct ways to form a $\Delta$-simplex from $m$ links.

At time $t$, there are $t$ nodes available for attachment, and the average degree is defined as $\bar{k}_t \equiv k_t/t$, where $k_t$ denotes the total degree of the network up to time $t$. At the MF level, we assume that all nodes have the \textit{same} degree $\bar{k}_t$, so that the probability of the $i$-th single attachment to a typical node is
\begin{equation}
	p_{\text{attachment}}^{(i)}=\frac{\bar{k}_t}{k_t^{(i)}}=\frac{1}{t-i}.
\end{equation}
Note that $\bar{k}_t \propto t^{1/2}$ for the BA model.

The probability that a $\Delta$-simplex is created after a new node $v_t$ is added, given that a $(\Delta-1)$-simplex already exists (with nodes $v_1, v_2, \ldots, v_{\Delta}$), follows a simple multiplicative rule. The probability increment $\delta P$ associated with $\Delta$ successive attachments to a $(\Delta-1)$-simplex is given by the product of the corresponding single-attachment probabilities, multiplied by the number of $(\Delta-1)$-simplices present at time $t-1$. This relation constitutes the MF master equation of the combinatorial theory:  
\begin{widetext}
\begin{equation}
		\delta P^{\text{MF}2}_\Delta(t,m) = \binom{m}{\Delta}\cdot\left[\prod_{i=0}^{\Delta-1}(\Delta-i)p_{\text{attachemnt}}^{(i)}\right] P^{\text{MF}2}_\Delta(t-1,m)=\frac{m!}{(m-\Delta)!}\frac{(t-(\Delta-2))!}{t!}\times P^{\text{MF}2}_\Delta(t-1).
\end{equation}
\end{widetext}
In this approximation, we assumed that the newly added node can attach to all the other existing nodes in the graph with the same probability. Using the Stirling's approximation~\cite{knopp1990theory} (see also the Appendix~\ref{SEC:Combinatorial}) for large enough times, we find
\begin{equation}
	\delta P^{\text{MF}2}_\Delta(t) =\frac{m!}{(m-\Delta)!}t^{-(\Delta-2)}	P^{\text{MF}2}_\Delta(t-1).
    \label{Eq:SigmaMF2}
\end{equation}
Using the fact that $ P^{\text{MF}2}_\Delta(t)\propto \Sigma_{\Delta}(t)$, and $\delta P^{\text{MF}2}_\Delta(t)\propto \delta\Sigma_\Delta(t)$, we find the following expression ($\Sigma^{\text{MF}2}_\Delta(t-1)\approx\Sigma^{\text{MF}2}_\Delta(t)$)
\begin{equation}
	\frac{\delta \Sigma^{\text{MF}2}_\Delta(t)}{\Sigma^{\text{MF}2}_\Delta(t)} =\frac{m!}{(m-\Delta)!}t^{-(\Delta-2)},
	\label{Eq:SigmaMFMain}
\end{equation}
the solution of which is 
\begin{equation}
	\left. \delta \Sigma^{\text{MF}2}_\Delta(t)\right|_{t\to\infty} =F_\Delta^{\text{MF}2}(m)t^{-\psi_{\text{MF}2}(\Delta)}\Theta(m-\Delta),
	\label{Eq:MFSigma}
\end{equation}
where  
\begin{equation}
	\psi_{\text{MF}2}(\Delta)=\Delta-2,
	\label{Eq:Exponents2}
\end{equation}
and also (see Appendix~\ref{SEC:Combinatorial} for the details)
\begin{equation}
	F_\Delta^{\text{MF}2}(m)\equiv \zeta\left(\psi_{\text{MF}2}(\Delta)\right)\frac{m!}{(m-\Delta)!}.
\end{equation}
In this relation, $\Theta$ denotes the Heaviside step function, which accounts for the constraint $m \geq m_{\Delta}^S \equiv \Delta$. This condition ensures that $m$ exceeds the threshold $\Delta$, below which $\delta\Sigma$ vanishes. This result can be understood using the LTTS property of BA networks. Specifically, simplices of dimension $\Delta > m_{\Delta}^S$ cannot be formed, since adding a new node with $m$ links to an existing graph allows only the creation of simplices of dimension $\Delta$ or lower.

For this reason, an $m$-BA network can contain simplices only in the range $0 \leq \Delta \leq m$, with $\Sigma_{\Delta > m}=0$ for higher dimensions. The exponent $\psi_{\text{MF}2}(\Delta)$ should be compared with that in Eq.~\eqref{Eq:Exponents1}. In the regime $m \gg \Delta$, we obtain
\begin{equation}
F_\Delta^{\text{MF}2}(m)\equiv
\zeta\!\left(\psi_{\text{MF}2}(\Delta)\right)(m-\Delta)^{\Delta}.
\label{Eq:MF-F}
\end{equation}
Care must be taken when using Eq.~\eqref{Eq:MF-F}, as it is not valid for values of $m$ close to $\Delta$. Beyond providing the exact values of the exponents, this relation indicates that $\delta\Sigma$ exhibits power-law behavior with respect to both $m$ and $t$.

\subsection{Simulation Results}
In this section, we present the simulation results. The behavior of $\delta\Sigma_{\Delta}(t,m)$ as a function of $t$ for the specific case $m=29$ is shown in Fig.~\ref{fig:sigma_F_psi}a, indicating a power-law decay at sufficiently long times. This observation confirms the MF prediction of power-law decay discussed above. The peak visible at early times in Fig.~\ref{fig:sigma_F_psi}a is transient: it disappears for larger system sizes (longer times) and arises from the small number of nodes available at early times to form a simplex.

A linear fit of the curves in the log--log plot at large times confirms the MF prediction in Eq.~\eqref{Eq:MFSigma}, motivating the following ansatz:
\begin{equation}
	\delta\Sigma^{\text{PL}}_{\Delta}(t,m)
	= F_\Delta(m)\, t^{-\psi(m,\Delta)}\, \Theta(m-\Delta),
	\label{eqt:sigma_delta_t_m}
\end{equation}
where $F_\Delta(m)$ is a proportionality constant and $\psi(m,\Delta)$ is the power-law exponent, which generally depends on both $m$ and $\Delta$. These quantities differ from their MF counterparts. The role of the threshold, represented by the step function in Eq.~\eqref{eqt:sigma_delta_t_m}, will be discussed later.
The initials ``PL" for power law has been employed to denote the fact that we are approximating the function as a power law and to distinguish it from the true function $\delta\Sigma_\Delta$. The exponent $\psi(m,\Delta)$ and the factor $F_\Delta(m)$ have been obtained using power-law fittings. The $F_\Delta(m)$ values were determined from the linear fits in log-log scale by setting $\log t = 0$, the result of which is reported in the Fig.~\ref{fig:sigma_F_psi}b. The exponents $\psi(m,\Delta)$ are reported in Fig.~\ref{fig:sigma_F_psi}c. Note that $\psi$ is independent of $m$, so that we have $\psi=\psi(\Delta)$. We notice also that $\delta\Sigma_0(t)=1$ and $\delta\Sigma_1(t)=m$ for all $t$ values (see Fig.~\ref{fig:sigma_F_psi}a).

Figure~\ref{fig:sigma_F_psi}b suggests a similar (power-law) structure as the MF prediction Eq.~\ref{Eq:MF-F} for $F$, with some deviations concerning the numerical values of the exponents. Inspired by this MF prediction, and our numerical results, we suggest the following power-law dependence: 
\begin{equation}\label{eq:F_m}
    F_\Delta(m) = a_F(m-b_F)^{c_F}.
\end{equation}
This relation is confirmed by the results shown in the inset of Fig.~\ref{fig:sigma_F_psi}b, the fitting parameters of which are shown in Table \ref{tab:fitting_params} for different $\Delta$ values. 

To determine any time dependence of $F_\Delta(m)$, we multiplied numerically both sides of Eq.~\eqref{eqt:sigma_delta_t_m} by $t^\psi$, using $\psi(\Delta)$ values obtained from the fitting of $\delta\Sigma_\Delta(t,m)$. The result presented in Fig.~\ref{fig:sigma_F_psi_2}b shows a very slight dependence on time, which might also be due to the uncertainty in determining the values of $\psi(\Delta)$. Note that the graph in Fig.~\ref{fig:sigma_F_psi_2}b is in the logarithmic scale.

An important observation concerns the $\Delta$-dependence of the exponent $\psi$, shown in Fig.~\ref{fig:sigma_F_psi_2}c. Equations~\eqref{Eq:Exponents1} and~\eqref{Eq:Exponents2} suggest that this exponent behaves as
\begin{equation}
\psi(\Delta)=
\begin{cases}
0, & \Delta<\Delta_{\text{LC}}\equiv 2,\\[4pt]
1+\text{(logarithmic correction)}, & \Delta=\Delta_{\text{LC}},\\[4pt]
\dfrac{\Delta}{2}, & \Delta_{\text{LC}}<\Delta\lesssim\Delta_{\text{CO}}\approx 5,\\[6pt]
\Delta-2, & \Delta_{\text{CO}}<\Delta.
\end{cases}
\label{Eq:exponents}
\end{equation}

\begin{figure}
	\includegraphics{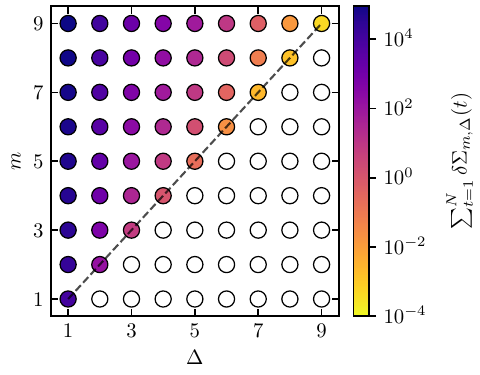}
	\caption{Simplicial phase transition: The total number of $\Delta$-dimensional simplexes at $t=N$ is shown for various combinations of $(m, \Delta)$. Colored disks represent parameter pairs where $\Delta$-simplexes are present, while empty disks correspond to $(m, \Delta)$ values for which such structures do not form. The boundary separating these regions, $m_{\Delta}^S=\Delta$, marks the critical parameters governing the emergence of higher-dimensional connectivity in the growing network.}
	\label{fig:m_delta_simplex}
\end{figure}

Here, LC stands for ``logarithmic correction,'' as obtained in Eq.~\eqref{Eq:SigmaMF1}. This behavior is reminiscent of the notion of an upper critical dimension in critical phenomena, where power-law scaling is modified by logarithmic factors. For the third branch ($\Delta_{\text{LC}}<\Delta\lesssim\Delta_{\text{CO}}$), we introduce a crossover point $\Delta_{\text{CO}}$ at which the system transitions from a low-fluctuation regime to a regime characterized by strong fluctuations. This crossover point is determined as the intersection between the $\Delta/2$ and $\Delta-2$ scaling forms. Due to the noise present in the numerical data—particularly in the regime $\Delta\to m$—it is difficult to verify these scaling relations for larger values of $m$ and $\Delta$. Figure~\ref{fig:m_delta_simplex} presents a color map of $\Sigma_\Delta(N,m)$ in the $(\Delta,m)$ parameter space, which clearly shows the threshold $m_\Delta^S=\Delta$ and illustrates how this quantity grows with increasing $m$ and $\Delta$.

\begin{figure*}
    \includegraphics[width = 0.95\textwidth]{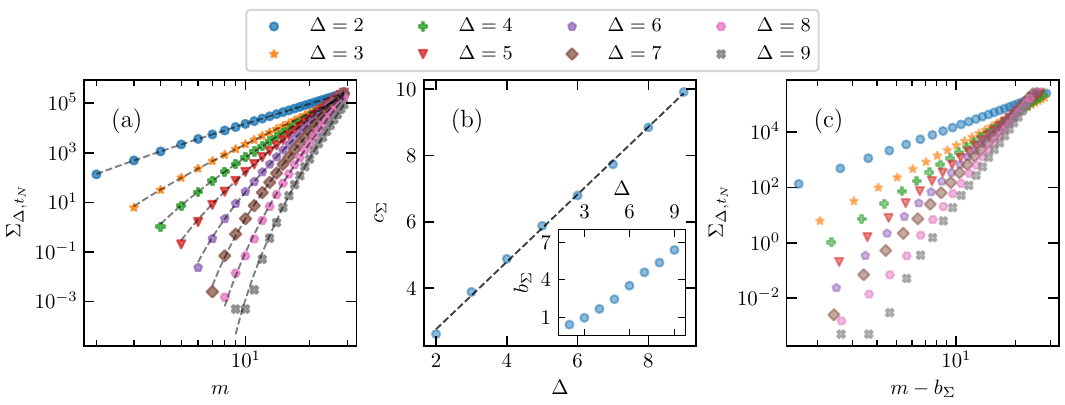}
    \caption{(a) The number of $\Delta$-dimensional simplexes at $t = t_{\mathrm{max}} = 10^4$ is shown for various combinations of $m$ and $\Delta$. The dashed lines represent the fitting curves described by Eq.~\eqref{eq:sigma}. (b) The corresponding fitting parameters, $b_{\Sigma}$ (inset) and $c_{\Sigma}$ (main panel), are plotted as functions of the simplex dimension $\Delta$. Error bars are smaller than the symbol size. The parameter $c_{\Sigma}$ exhibits a linear dependence on $\Delta$, with a slope of $1.02 \pm 0.02$ and an intercept of $0.72 \pm 0.09$. (c) When the number of $\Delta$-dimensional simplexes is plotted against $m - b_{\Sigma}$ in log–log scale, the data collapse onto straight lines, confirming the power-law behavior and supporting the validity of the proposed fitting function.}
    \label{fig:sigma_m}
\end{figure*}

At this point, we can make some parallels with centrality properties of networks such as the clustering coefficient $C$, which corresponds to the simple but nontrivial case where $\Delta=2$. Both, $\delta\Sigma_{2}$ and the concept of the clustering coefficient, especially the global clustering coefficient, target counting the number of triangles within the network. However, they are not equivalent. In the former case, we strictly count the number of triangles while in the latter case triangle counting is only part of the process. Nevertheless, it is interesting that the exponent $\psi_2(m)$ in Eq.~(\ref{eqt:sigma_delta_t_m}) has an apparent  value close to $3/4$ (Fig.~\ref{fig:sigma_F_psi}c), which is similar to the scaling $C \sim N^{-3/4}$ obtained for the clustering coefficient of the BA model~\cite{albert2002statistical}. As this exponent $3/4$ is a numerical artifact of the ``logarithmic correction" in our case, we think that a similar care is also needed for the clustering coefficient of BA.\\ 

Due to the LTTS property, we expect that $\Sigma_\Delta(N,m)$ increases monotonically with $m$ for all $\Delta$ values, implying that the probability of having $\Delta$-simplexes increases with $m$. This is confirmed by the numerical results shown in Figs.~\ref{fig:m_delta_simplex} and~\ref{fig:sigma_m}a. Moreover, from Eq.~\ref{eqt:sigma_delta_t_m}, one expects the follwoing form for $\Sigma_\Delta(N, m)$, 
\begin{equation}\label{eq:sigma}
    \Sigma_\Delta(N, m) = a_\Sigma(m-b_\Sigma)^{c_\Sigma},
\end{equation}
where $a_\Sigma$, $b_\Sigma$ and $c_\Sigma$ is a new set of fitting parameters. This relation is confirmed by the analysis presented in Figs.~\ref{fig:sigma_m}b and~\ref{fig:sigma_m}c, where the log-log plots of $\Sigma_\Delta(N, m)$ in terms of $m-b_\Sigma$ are linear. Note that, in case of pure power-law behavior according to Eqs.~\ref{eqt:sigma_delta_t_m} and~\ref{eq:F_m}, these parameters should be the same as the fitting set in Eq.~\ref{eq:F_m}. Given that the time dependence of $\delta\Sigma$ is not pure power-law, some deviations are expected, which is why we have used a new set of parameters. The results of the fitting can be found in Table \ref{tab:fitting_params}, confirming that there are differences between the set of exponents here and the ones in Eq.~\ref{eq:F_m} (more details can be found in the appendix).\\

The power-law behavior in the figure points towards a phase transition with respect to the parameter $m$. Below $m=b_\Sigma$, the total number of $\Delta$-simplexes is zero, while above this transition value we see a power-law. An important fact is that $\Sigma_\Delta(m)$ is not a continuous curve. It presents a finite jump at the transition point, which decreases for increasing $\Delta$. The size of the jump $G_\Delta$, or discontinuity, can be obtained from Eq.~\eqref{eq:sigma}, by substituting $m=m_\Delta^S=\Delta$ to obtain
\begin{equation}
    G_\Delta \equiv \Sigma_\Delta(N, m=\Delta) = a_\Sigma(\Delta-b_\Sigma)^{c_\Sigma}.
\end{equation}
It is this jump that accounts for the discrepancy between the fitted parameter $b_\Sigma$ and $m_\Delta^S=\Delta$.  In Table \ref{tab:fitting_params} we can see that $b_\Sigma$ is smaller than $\Delta$ for all values of $\Delta$ considered. However, it also shows that the value of $b_\Sigma$ gets closer to $\Delta$ as the latter increases. This behavior is in line with the fact that the gaps in $\Sigma_\Delta$ decreases with increasing $\Delta$ as seen in Fig.~\ref{fig:sigma_m}c.

\begin{figure}
    \includegraphics{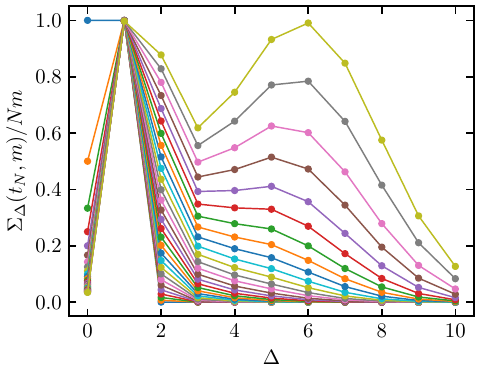}
    \caption{The number of $\Delta$-dimensional simplexes at  $t=t_{max}=10^4$ for different $m$ and $\Delta$ values normalized by $Nm$. The range of $m$ is from 1 (the line at the bottom, for $\Delta>1$) to 29 (the line at the top, for $\Delta>1$).}
    \label{fig:total_number_of_simplex}
\end{figure}

In Fig.~\ref{fig:total_number_of_simplex}, we present the $\Delta$-dependence of a normalized $\Sigma$ for different connectivity values $m$. The figure presents two peaks (ignoring $\Delta=0$), one at $\Delta=1$ for all $m$ and the other that moves towards larger $\Delta$ values for increasing $m$. While the former is trivial, because of the total number of 1-simplexes being equal to $Nm$ for all $m$'s,  the latter is more complex in nature. 

Following Eq.~\eqref{eqt:sigma_delta_t_m} and therefore approximating $\delta\Sigma$ as a pure power-law, the final number of $\Delta$-simplexes contained in the graph $G$ for a chosen $m$, can be written as 
\begin{equation}
    \begin{split}
        \Sigma^{\text{PL}}_{\Delta}(N,m) = F_\Delta(m)H(N,\psi(\Delta))\Theta(m-\Delta),
    \end{split}
 \label{eq:total_simplexes_pl}
\end{equation}
where $H(t,r)$ is given in Eq.~\ref{Eq:HarmonicNumber}, and PL again reminds the fact that the sum is associated with the approximation in Eq.~\eqref{eqt:sigma_delta_t_m}, valid only in the power-law regime. By using Eq.~\eqref{eq:total_simplexes_pl}, we have that for $m > m_\Delta^S$, $\Sigma^{\text{PL}}_\Delta(N,m) = F_\Delta(m)H_\Delta(N)$ and therefore $\Sigma^{\text{PL}}_\Delta(N,m) \propto F_\Delta(m)$. 

We tried to determine how well $\delta\Sigma^{PL}_\Delta$ approximates $\delta\Sigma_\Delta$ by determining how well the proportionality $\Sigma\Delta \propto F_\Delta$ holds. We used the results of our simulations and summed the values of $\delta\Sigma_\Delta$ cumulatively to determine directly $\Sigma_\Delta(m)$. On the other hand, as described earlier, the values of $F$ correspond to one of the two fitting parameters for the relationship between $\delta\Sigma^{PL}$ and time. Details of the fitting can be found in the appendix. We used the values of $F$ and $\Sigma$ for different combinations of $\Delta$ and $m$ to fit the functions \ref{eq:F_m} and \ref{eq:sigma} respectively. The proportionality of the two functions and therefore the equivalence between $\Sigma_\Delta$ and $\Sigma^{PL}_\Delta$ requires that $b_F = b_\Sigma$ and $c_F = c_\Sigma$, within statistical error. 

The results of the fitting processes for $F_\Delta(m)$ and $\Sigma_\Delta(m)$ are shown in Table~\ref{tab:fitting_params} and Fig.~\ref{fig:sigma_m}b. While there is significant statistical overlap between some of the fitting parameters $b_F/b_\Sigma$ and $c_F/c_\Sigma$, in most cases there is clearly no overlap. Therefore, we cannot say that $\Sigma_\Delta = \Sigma^{\text{PL}}_\Delta$. Nevertheless, it does show that $\Sigma^{PL}_\Delta$ approximates $\Sigma_\Delta$ remarkably well.

\begin{table*}[!ht]
\caption{The fitting parameters obtained for the $m$-dependence of $F$ and $\Sigma$ for different values of the dimension $\Delta$. We have used the models $F(m) = a_F(m-b_F)^{c_F}$ and $\Sigma(m) = a_\Sigma(m-b_\Sigma)^{c_\Sigma}$. More details can be found in the appendix.}
\begin{ruledtabular}
\begin{tabular}{ccccc}
parameters    & $\Delta=2$        & $\Delta=3$      & $\Delta=4$    & $\Delta=5$ \\ \hline
$b_{F}$       & $0.597\pm0.008$   & $1.05\pm0.02$   & $1.60\pm0.06$   & $2.0\pm0.2$ \\ 
$b_{\Sigma}$ & $0.391\pm0.001$   & $0.942\pm0.004$ & $1.653\pm0.009$ & $2.43\pm0.02$  \\
$c_{F}$      & $2.419\pm0.002$   & $4.129\pm0.006$ & $5.45\pm0.02$   & $7.06\pm0.09$ \\
$c_{\Sigma}$ & $2.6089\pm0.0002$ & $3.882\pm0.001$ & $4.882\pm0.003$ & $5.887\pm0.009$\\ 
\end{tabular}
\end{ruledtabular}
\label{tab:fitting_params}
\end{table*}

\section{Graph Homology and Betti number statistics}\label{sec:topology}

In this section, we study the homological properties of BA networks by computing the dimension of the $\Delta^{th}$ homology group of clique simplicial complex $\mathcal{K}(G)$ mapped from the network $G$. This topological invariant, called the $\Delta^{th}$ \textit{Betti number} $\beta_{\Delta}$, counts $\Delta$-dimensional topological holes ($\Delta$-holes) of the topological space underlying the complex network. A $\Delta$-hole of a space (network) is a subspace (sub-network) which has no boundary (a $\Delta$-cycle) and is also not a $\Delta$-boundary of any subspace of the original space. Therefore, $\beta_{0}$ counts the number of connected components of the network, $\beta_{1}$ the number of topological loop structures (related with the basic cycles) of the network, $\beta_{2}$ the number of topological voids and so on. 

We employ the Dionysus Python library~\cite{dmitriydionysus} to compute the time-dependent persistent homology of the BA model using a Vietoris--Rips filtration that begins with $N$ isolated vertices. As time evolves and edges from the BA model are introduced, we construct a sequence of nested Vietoris--Rips complexes indexed by time, continuing until the full BA network is recovered. The sole difference between the BA network and the complexes generated through the Vietoris--Rips filtration concerns the number of connected components ($\beta_0$). In the BA model, $\beta_0$ is always equal to one, whereas in the Vietoris--Rips filtration it starts at $N$, decreases by one at each step, and eventually reaches one. Because the zero-dimensional homology is trivial in this setting, we do not examine the connected components of the BA model. Instead, we focus on the higher-dimensional, nontrivial topological features, which coincide for both the BA network and its Vietoris--Rips filtration.

\begin{figure*}
    \subfigure[]{\includegraphics[width=0.3\textwidth]{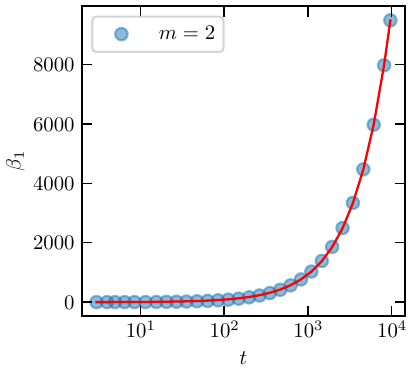}}
    \subfigure[]{\includegraphics[width=0.3\textwidth]{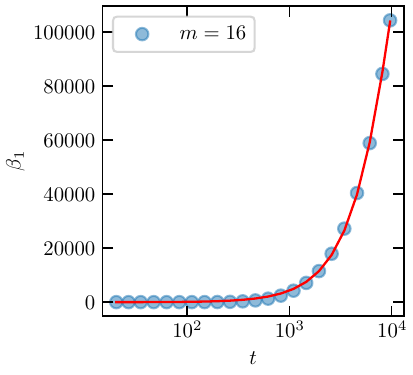}}
    \subfigure[]{\includegraphics[width=0.3\textwidth]{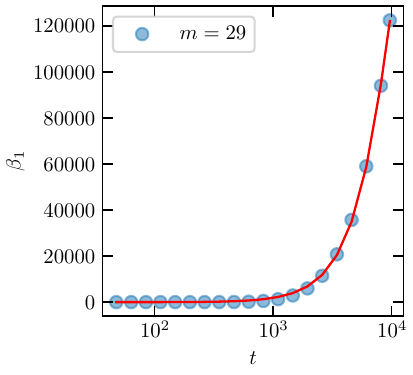}}
    	\subfigure[]{\includegraphics[width=0.3\textwidth]{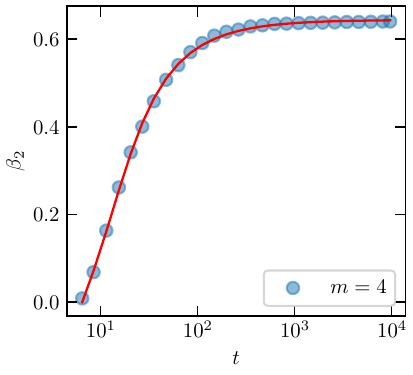}}
	\subfigure[]{\includegraphics[width=0.3\textwidth]{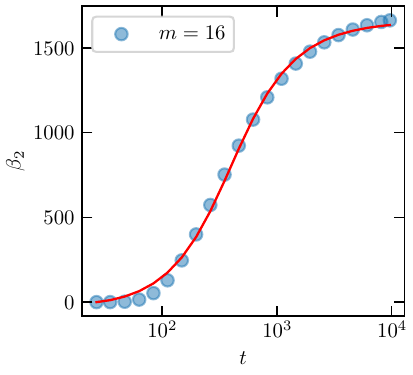}}
    	\subfigure[]{\includegraphics[width=0.3\textwidth]{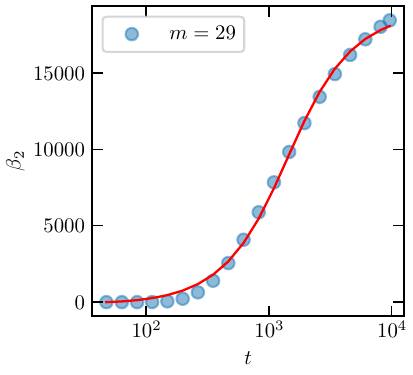}}
    \subfigure[]{\includegraphics[width=0.3\textwidth]{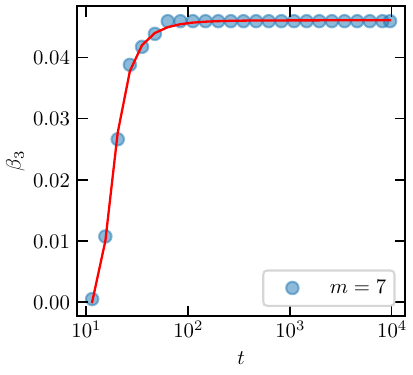}}
    \subfigure[]{\includegraphics[width=0.3\textwidth]{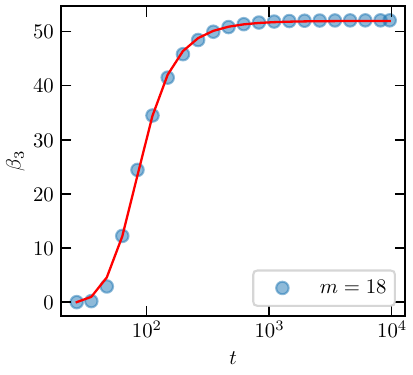}}
    \subfigure[]{\includegraphics[width=0.3\textwidth]{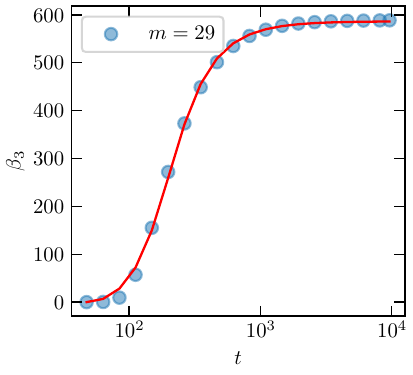}}
    \subfigure[]{\includegraphics[width=0.3\textwidth]{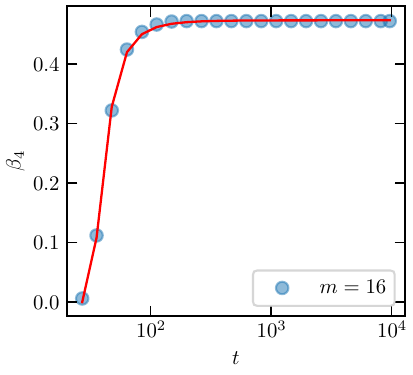}}
    \subfigure[]{\includegraphics[width=0.3\textwidth]{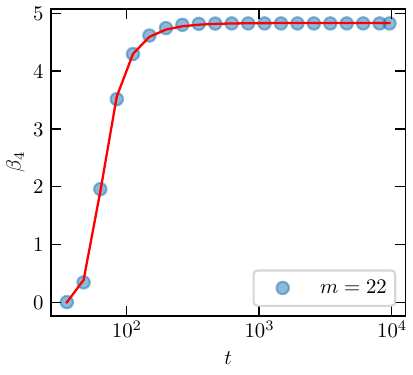}}
    \subfigure[]{\includegraphics[width=0.3\textwidth]{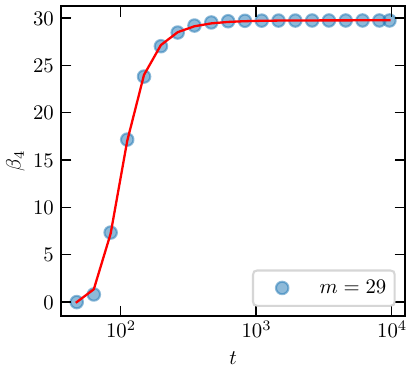}}
	\caption{Time-dependent variation of $\beta_1$  for different $m$ values: (a) 2, (b) 16, and and (c) 29. (d), (e) and (f) show $\beta_2$ as a function of $t$ for $m$ values of 4, 16 and 29. (g), (h) and (i) show $\beta_3$ as a function of $t$ for $m$ values of 7, 18 and 29. (j), (k) and (l) show $\beta_4$ as a function of $t$ for $m$ values of 16, 22 and 29. The solid lines show the data fittings by the Eq.~\eqref{Eq:betti_vs_t}.}
	\label{fig:beta_vs_t_dim1}
\end{figure*}

\begin{figure}
	\includegraphics{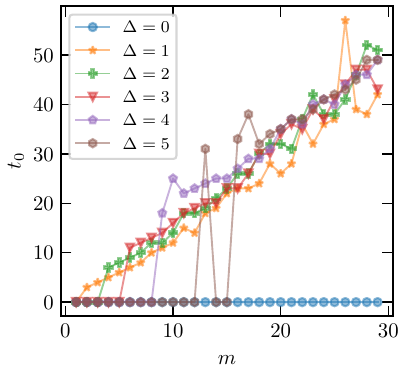}
	\caption{The initial time step, $t_0$, marking the first appearance of a $\Delta$-dimensional hole during the network evolution, is presented as a function of both the connectivity parameter $m$ and the dimension $\Delta$.}
	\label{fig:t0_m}
\end{figure}

\begin{figure*}
    \includegraphics[width=\textwidth]{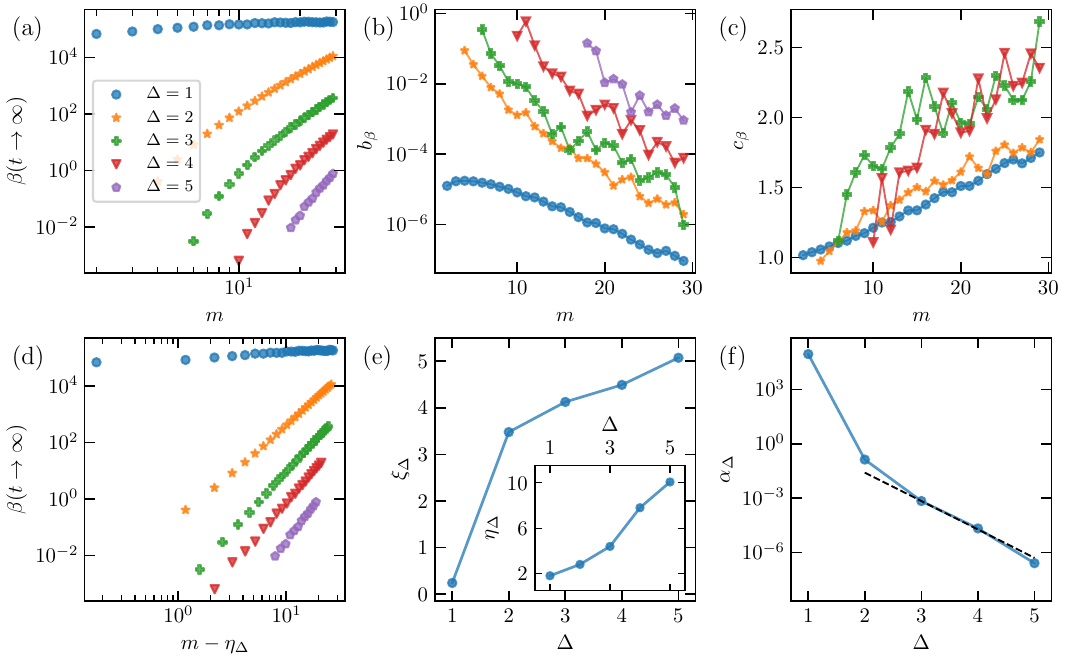}
	\caption{The evolution of the $\Delta$-dimensional Betti numbers for different $m$ values is fitted using Eq.~\eqref{Eq:betti_vs_t} as a function of time $t$. Panels (a), (b), and (c) display the variation of the fitting parameters $\beta(t \to \infty)$, $b_{\beta}$, and $c_{\beta}$, respectively. The data in panel (a) are further fitted by Eq.~\eqref{eq_betti_infinity_fitting}, and as shown in panel (d), these data become linear in log–log scale when plotted against $m - \eta_{\Delta}$. Panel (e) presents the dependence of the parameters $\eta_{\Delta}$ (main panel) and $\xi_{\Delta}$ (inset) on the simplex dimension $\Delta$. Finally, panel (f) shows that the parameter $\alpha_{\Delta}$, defined in Eq.~\eqref{eq_betti_inf_alpha}, exhibits an exponential decay for $\Delta > 1$, while the dashed line shows the fitting. The error bars in (e) and (f) are smaller than the symbol sizes.}
	\label{fig:betti_parameters}
\end{figure*}

By growing the BA network, topological features evolve and change in different ways. Analogous to the case of simplicial properties, the existence of $\Delta$-holes within the network depends directly on the connectivity parameter $m$. We also find in this case that there exists a minimum connection number, denoted by $m_{\Delta}^H$, that ensures the possible existence of $\Delta$-dimensional holes in the network. It is important to note that in the case of homological analysis of BA model, the existing topological features may be altered after adding a new node with $m$ links. Therefore, the addition of a new node can destroy an existing hole in the graph. An important question that we want to address is how the system behaves beyond the threshold $m_{\Delta}^H$. Tackling this problem reveals that the system exhibits a rich behavior in terms of $m$ and $\Delta$. We used the simulations described in the previous section to determine the evolution of the Betti numbers up to dimension six and consequently their dependence on time and the connectivity number $m$. Figure~\ref{fig:beta_vs_t_dim1} shows the behavior of Betti numbers as a function  of time for different dimensions and $m$ values. We examined various consistent trial functions as the possible fitting of the $\beta_{m,\Delta}-t$ relation. This list includes, but is not restricted to power-law, modified exponential, logarithmic, and log-normal, and ArcTan decay. By extracting the goodness-of-fitting parameters ($R^2$, confidence interval, AIC, and BIC), we realized that, among a long list of consistent functions, the best fitting arises using the ArcTan decay, as is explained in the remainder. This fitting function reads
\begin{equation}
    \beta_{m,\Delta}(t)=\frac{2\beta(t\to\infty)}{\pi}\tan^{-1}\left[b_{\beta}\left(t-t_0^{(\beta)}\right)^{c_{\beta}}\right]_+,
    \label{Eq:betti_vs_t}
\end{equation}
where the parameters $\beta(t\to\infty)$, $b_{\beta}$, $c_{\beta}$, and $t_0^{(\beta)}$ depend on $\Delta$ and $m$, and $[f(t)]_+$ is $f(t)$ for $t>t_0^{(\beta)}$ values, and is zero for $t\le t_0^{(\beta)}$. The parameter $t_0^{(\beta)}$ specifies the time step at which the topological holes appear for the first time in the dynamics of the BA model, which is represented in Fig.~\ref{fig:t0_m}. In this equation $b_{\beta}$ is the inverse decay time for the inverse tangent function, $c_{\beta}$ is the growth exponent, and $\beta_\Delta(t\to\infty)$ is the Betti number in the long time limit which the betti functions saturate to. Note that, for $\delta t\equiv \frac{t-t_0^{(\beta)}}{t_0^{(\beta)}}\ll 1$, this function is transformed to a power-law relation:
\begin{equation}
\left. \beta_{m,\Delta}(t)\right|_{0<\delta t\ll 1}\propto \left(t-t_0^{(\beta)}\right)^{c_{\beta}},
\end{equation}
which is a critical dynamical contentious transition, endowed with a power-law behavior. This tells us that $c_{\beta}$ is actually a critical exponent. This function is analyzed in Fig.~\ref{fig:betti_parameters}. The numerical results suggest that $\beta(t\to\infty)$ function shows also a power-law behavior in terms of $m$ as follows:
\begin{equation}
\beta_{m,\Delta}(t\to\infty)=\alpha_\Delta(m-\eta_\Delta)^{\xi_\Delta}\Theta(m-m_{\Delta}^H)
\label{eq_betti_infinity_fitting}
\end{equation}
where $\alpha_\Delta$, $\eta_\Delta$, $\xi_\Delta$, and $m_\Delta^H$ are fitting parameters. $\xi_\Delta$ is a scaling exponent, and $m_\Delta^H$ is a threshold (topological transition point) below which there is no $\Delta$-hole. $\alpha_\Delta$ decays exponentially with $\Delta$ as follows (Fig.~\ref{fig:betti_parameters}f):
\begin{equation}
    \alpha_\Delta \approx Ae^{-\Delta/\Delta_0} 
    \label{eq_betti_inf_alpha}
\end{equation}
for $\Delta>1$, and $\Delta_0=0.277\pm 0.001$. The fact that $ m_\Delta^H - \eta_\Delta\ge 0$ shows that the transition is gapful in the sense that there is a finite jump at the transition point $m_\Delta^H$. The dependency of $b_{\beta}$ and $c_{\beta}$ on $\Delta$ and $m$ is shown in the Figs.~\ref{fig:betti_parameters}b and~\ref{fig:betti_parameters}c, respectively. Note that $b_{\beta}$ decays exponentially with $m$, while $c_\beta$ grows almost linearly with $m$. Both these observations show that the saturation to the final value of inverse tangent is slower for larger values of $m$.\\ 

The Eq.~\eqref{Eq:betti_vs_t} can be viewed as a dynamical equation:
\begin{equation}
\dot{\tilde{\beta}}=\frac{1}{\pi}c_\beta b_\beta^{\frac{1}{c_\beta}}\sin\pi\tilde{\beta}\left(\cot \frac{\pi\tilde{\beta}}{2}\right)^{\frac{1}{c_\beta}},
\label{Eq:dynamicbeta}
\end{equation}
where $\tilde{\beta}\equiv\frac{\beta(t)}{\beta(t=\infty)}$. This relation tells us that there are two important fixed points of the dynamics: $\tilde{\beta}=0,1$ (note that, this relation predicts also other type of fixed points, irrelevant to our discussion), the former being the starting point of the dynamics, which is always unstable towards the latter being the point at which the dynamic stops asymptotically $\tilde{\beta}=1$ or $\beta=\beta(t\to\infty)$. For the $\beta=0$ fixed point to be non-singular and unstable towards the other, one can expand the right hand side of Eq.~\ref{Eq:dynamicbeta} around $\beta=0$, which tells us that
\begin{equation}
\dot{\tilde{\beta}}(\tilde{\beta}\ll 1)\approx \frac{c_\beta(2b_\beta)^{1/c_\beta}}{\pi^{1+1/c_\beta}}\tilde{\beta}^{1-1/c_\beta},
\end{equation}
 which implies that $c_\beta\ge 1$ for all the cases. We see from Fig.~\ref{fig:betti_parameters}c that this requirement is always satisfied. 

\begin{figure}
	\includegraphics{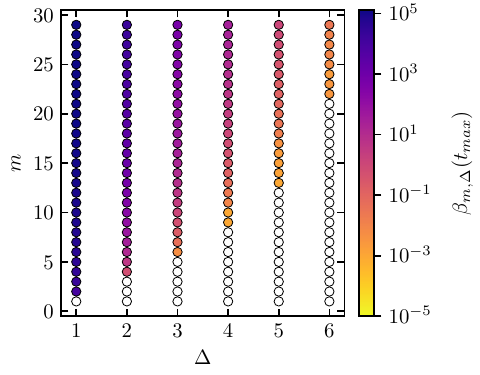}
	\caption{Homological phase transition: The total number of $\Delta$-dimensional holes, $\beta_{m,\Delta}(t_{\mathrm{max}})$, is shown for various combinations of $(m, \Delta)$, revealing a distinct phase transition. Colored disks indicate parameter pairs where nonzero Betti numbers are observed, while empty disks correspond to $(m, \Delta)$ values for which no holes are present. The emerging boundary between these regions delineates the critical set of parameters associated with the onset of topological transitions.}
	\label{fig:total_number_of_hole}
\end{figure}

Figure~\ref{fig:total_number_of_hole} shows our numerical results for the dimension-dependency of the connectivity number. It explicitly demonstrates that a network cannot contain topological holes of dimension $\Delta$ if the connectivity parameter $m$ is less than $m_{\Delta}^H$. Once $m\geq m_{\Delta}^H$, it is necessary to wait on average $t_0$ time steps for the first $\Delta$-dimensional hole of the BA model to be created.

\section{Concluding Remarks}~\label{SEC:conclusion}
In this study, we examined the temporal evolution of simplicial complexes derived from the BA model, focusing on the scaling and critical behaviors of $\Delta$-dimensional simplexes and holes as the network evolves in time. By employing the filtration parameter $t$ to represent network growth, we investigated how the higher-order topological structure develops and transforms as new edges are added. Our analysis of Betti numbers up to dimension six provides a comprehensive characterization of these transformations and elucidates how a topological complexity emerges from the underlying preferential-attachment mechanism.

For the growth of the simplicial complexes, we developed two different MF theories, each working in a distinct regime. We found an analytic form of the growth function of the $\Delta$-simplexes showing scaling behavior in the long time regime with the exponents according to Eq.~\ref{Eq:exponents}. The results demonstrate that the number of $\Delta$-dimensional simplexes follows robust scaling laws with respect to time, Figs.~\ref{fig:sigma_F_psi} and~\ref{fig:sigma_F_psi_2}, suggesting the existence of self-similar topological growth patterns inherent to the BA model. Equations~\ref{eqt:sigma_delta_t_m} and~\ref{eq:F_m}, and~\ref{eq:sigma} describe the dynamical aspects of the simplicial complexes, which are in accordance with the MF theories  developed in this paper. This scaling behavior reflects the complex organization of the network’s higher-order structures and highlights the fact that preferential attachment not only shapes the power-law distribution function of statistical quantities, like the degree and betweenness~\cite{barabasi1999emergence}, but also shapes complex (scaling) topological observables, defined in this study. The $\Delta$-dimensional holes as the homological observables exhibit a nontrivial dependence on time with a threshold, well captured by an arctangent functional form; refer to Fig.~\ref{fig:beta_vs_t_dim1}. Equations~\ref{Eq:betti_vs_t},~\ref{eq_betti_infinity_fitting}, and~\ref{eq_betti_inf_alpha} deals with the dynamical properties of $\Delta$-holes in terms of time. This arctangent trend indicates the presence of saturation phenomena, where topological features proliferate rapidly in early growth stages but reach asymptotic stability as the network becomes denser and more interconnected.

A particularly noteworthy outcome of our analysis is the observation of topological transitions (TT) at some thresholds, which depend on $\Delta$ and $m$, the configuration of which are represented in Figs.~\ref{fig:m_delta_simplex} and~\ref{fig:total_number_of_hole}, which delineate regimes with different topological organization. Such critical behavior underscores the analogy between topological evolution in growing networks and phase transitions in statistical physics. Moreover, the dependence of these transitions on the connectivity parameter $m$ emphasizes that the number of links each new node introduces plays a pivotal role in determining not only local connectivity but also the global topological phase of the system.

\section*{Author contributions}

V. Adami: data analysis, figures and manuscript preparation; H. Masoomy: data preparation, data analysis and simulations; M. Lukovi\'c: simulations, data interpretation and manuscript preparation; M. N. Najafi: research idea, research conceptualization, data interpretation and manuscript preparation.\\

\section*{Acknowledgement}

\appendix

\section{Wick's theorem for the Probability of Creation of a $\Delta$-simplex}\label{SEC:Wicks}
Here we prove Eq.~\ref{Eq:ProbSigma}. Using Eq.~\ref{Eq:prob_ij}, we find
\begin{widetext}
	\begin{equation}
		\begin{split}
			P_\Delta(t)&\propto\sum_{t_{i_1}=1}^t\sum_{t_{i_2}\ne t_{i_1}}^t\sum_{t_{i_3}\ne t_{i_2}\ne t_{i_2}}^t...\sum_{t_{i_{\Delta+1}}\ne t_{i_1}...\ne t_{\Delta}}^t\prod_{\left\lbrace m,n \right\rbrace } \frac{m}{2\sqrt{t_mt_n}}\\
			&=\left(\frac{m}{2}\right)^{\frac{1}{2}\Delta(\Delta+1)}\sum_{t_{i_1}=1}^t\sum_{t_{i_2}\ne t_{i_1}}^t\sum_{t_{i_3}\ne t_{i_2}\ne t_{i_2}}^t...\sum_{t_{i_{\Delta+1}}\ne t_{i_1}...\ne t_{\Delta}}^t\frac{1}{t_{i_1}^{\frac{\Delta}{2}}}\frac{1}{t_{i_2}^{\frac{\Delta}{2}}}\frac{1}{t_{i_3}^{\frac{\Delta}{2}}}...\frac{1}{t_{i_{\Delta+1}}^{\frac{\Delta}{2}}}\\
			&=\left(\frac{m}{2}\right)^{\frac{1}{2}\Delta(\Delta+1)}\sum_{t_{i_1},t_{i_2},t_{i_3},...,t_{i_{\Delta+1}}=1}^t\left(t_{i_1}t_{i_2}t_{i_3}...t_{i_{\Delta+1}}\right)^{-\frac{\Delta}{2}}\prod_{i, j=1}^{\Delta+1}\left(1-\delta_{t_i,t_j}\right)
		\end{split}
		\label{Eq:ProbSigma2}
	\end{equation}
\end{widetext}
In the second line, we used the fact that each $\sqrt{t_i}$ repeats $\Delta$ times in each simplex, and the last line contains a product of $1-\delta_{t_i,t_j}$, where $\delta$ is a Kronecker delta function. This function can be expanded in terms of the times. Important examples are
\begin{equation}
\begin{split}
&\text{for}\ \Delta=1; \\
&\text{the factor is}\ 1-\delta_{t_1,t_2}\\
&\text{for}\ \Delta=2; \\
&\left(1-\delta_{t_1,t_2}\right)\left(1-\delta_{t_2,t_3}\right)\left(1-\delta_{t_1,t_3}\right)=1-\delta_{t_1t_2}-\delta_{t_2t_3}-\delta_{t_1t_3}\\
&+\delta_{t_1t_2}\delta_{t_2t_3}+\delta_{t_1t_3}\delta_{t_3t_2}+\delta_{t_2t_1}\delta_{t_1t_3}-\delta_{t_1t_2}\delta_{t_2t_3}\delta_{t_1t_3}.	
\end{split}
\end{equation}
This relation for the general case is reminiscent of the ``contraction" in the well-known Wick's theorem. Using the formal definition of the harmonic number 
\begin{equation}
H(t,r)=\sum_{t_i=1}^tt_i^{-r},
\end{equation}
one can find the expression for each term. It is instructive to find these terms for smallest $\Delta$ values. Dropping the pre-factor $\left(\frac{m}{2}\right)^{\frac{1}{2}\Delta(\Delta+1)}$, one finds
\begin{equation}
\begin{split}
&\text{for}\ \Delta=1;\\
&P_1(t)\propto H\left(t,\frac{1}{2}\right)^2-H(t,1),\\
&\text{for}\ \Delta=2;\\
&P_2(t)\propto H(t,1)^3-3H(t,2)H(t,1)+2H\left(t,\frac{3}{2}\right),\\
\end{split}
\label{Eq:sigma1,2}
\end{equation}
Doing the expansion for $\Delta=3$ is more challenging. For this case
\begin{itemize}
	\item There is one contraction configuration associated with the factor $1$ (zero $\delta$ functions), for which the summation is $H\left(t,\frac{3}{2}\right)^4$,
	\item There are $6$ contraction configurations associated with one $\delta$ function, for which the summation is $-6H(t,3)H\left(t,\frac{3}{2}\right)^2$,
	\item There are $15$ contraction configurations associated with two $\delta$ functions, (12 of which there is one isolated node, and 3 with non-isolated nodes) for which the summation is $12H\left(t,\frac{9}{2}\right)H\left(\frac{3}{2}\right)+3H(t,3)^2$,
	\item There are $40$ contraction configurations associated with three $\delta$ functions, (4 of which there is one isolated node, and 36 with non-isolated nodes) for which the summation is $-4H\left(t,\frac{9}{2}\right)H\left(\frac{3}{2}\right)-36H(t,6)$,
	\item There are $15$ contraction configurations associated with four $\delta$ functions, for which the summation is $15H\left(t,6\right)$,
	\item There are $6$ contraction configurations associated with five $\delta$ functions, for which the summation is $-6H\left(t,6\right)$,
	\item There are $1$ contraction configuration associated with six $\delta$ functions, for which the summation is $H\left(t,6\right)$.
\end{itemize}
The result is then
\begin{widetext}
\begin{equation}
P_3(t)=H\left(t,\frac{3}{2}\right)^4-6H(t,3)H\left(t,\frac{3}{2}\right)^2+8H\left(t,\frac{9}{2}\right)H\left(t,\frac{3}{2}\right)+3H(t,3)^2-26H(t,6)
\label{Eq:sigma3}
\end{equation} 
\end{widetext}
This method can be continued to higher $\Delta$ values. \\

Now we concentrate on the asymptotic behavior of $\delta\Sigma_\Delta(t)\propto \partial_tP_\Delta(t)$ in the limit $t\to\infty$. Using the expressions found above we have
\begin{equation}
\left. \delta\Sigma_1(t) \right|_{t\to\infty}\propto 4+\zeta\left(\frac{1}{2}\right)t^{-\frac{1}{2}}-\frac{1}{t}+O\left(t^{-1}\right)
\end{equation}
\begin{equation}
\left. \delta\Sigma_2(t) \right|_{t\to\infty}\propto 3\frac{\left(\ln t\right)^2}{t}-\frac{\pi^2}{2t}-3\frac{\left(\ln t\right)^2}{2t^2}+2t^{-\frac{3}{2}}-3\frac{\ln t}{t^2}+O(t^{-\frac{5}{2}})
\end{equation}
\begin{widetext}
\begin{equation}
\begin{split}
\left. \delta\Sigma_3(t) \right|_{t\to\infty}\propto &\frac{4\zeta\left(\frac{3}{2}\right)^3+6\psi^{(2)}(1)\zeta\left(\frac{3}{2}\right)+8\zeta\left(\frac{9}{2}\right)}{t^{\frac{3}{2}}}-12\frac{2\zeta\left(\frac{3}{2}\right)^2+\psi^{(n)}(1)}{t^2}+3\frac{\zeta\left(\frac{3}{2}\right)\left(16-\zeta\left(\frac{3}{2}\right)^2\right)-\frac{3}{2}\psi^{(2)}(1)\zeta\left(\frac{3}{2}\right)-2\zeta\left(\frac{9}{2}\right)}{t^{\frac{5}{2}}}\\
&+O(t^{-3}),
\end{split}
\end{equation}
\end{widetext}
where $\zeta(z)$ is the Riemann zeta function, and $\psi^{(n)}(z)$ is the $n$th derivative of digamma function. The first term in all the cases (Eqs.~\ref{Eq:sigma1,2} and Eq.~\ref{Eq:sigma3}) give the correct scaling behavior in long enough times:
\begin{equation}
\left. P_{\Delta}(t)\right|_{\text{first term}}= H\left(t,\frac{\Delta}{2}\right)^{\Delta+1}.
\label{Eq:approximation}
\end{equation}
This term - which corresponds to the case where there are no restrictions in the summations in Eq.~\ref{Eq:ProbSigma2} - gives asymptotically the correct scaling exponent of the model, and is valid for not large $\Delta$ values. Note that only the scaling behavior of $P_\Delta(t)$ is asymptotically expected to be described by this first term. One may then be interested in the asymptotic behavior of the Shannon entropy of the $\Sigma$ objects, in the MF level, which is given by
\begin{equation}
\mathcal{S}(t)=\sum_\Delta P_\Delta(t)\ln P_\Delta(t),
\end{equation}
for which an exact asymptotic expression is found using Eq.~\ref{Eq:approximation}.\\

The asymptotic behavior of $\delta\Sigma$ in the MF level is then given by
\begin{equation}
\delta\Sigma_\Delta(t\to\infty)\propto\left\lbrace \begin{matrix}
t^{-\psi_{\text{MF}1}(\Delta)},\ \ \Delta\ne 2,\\
t^{-1}\left(\ln t\right)^2,\ \ \Delta=2,
\end{matrix}\right. 
\end{equation}
where 
\begin{equation}
\psi_{\text{MF}1}(\Delta)=\left\lbrace \begin{matrix}
0 & \text{for}\ \Delta=0,1,\\
\frac{\Delta}{2} & \text{for}\ \Delta\ge 3.
\end{matrix}\right. 
\end{equation}
For large $\Delta$ values ($6\lesssim \Delta$), the exponent is different from the numerical values, showing that the approximation Eq.~\ref{Eq:approximation} is not reliable. The reason is the large fluctuations of the model for large $\Delta$ values. In the next section, we develop another MF theory, which matched better with the results in this regime.

\section{MF theory for $\Delta$-simplex; A Combinatorial Point of View}\label{SEC:Combinatorial}

In this section, we present an alternative MF theory appropriate for the case in which fluctuations in attachments are higher than usual (the case that led to inconsistencies in the previous section). In this regime, we introduce a master equation governing the probability of attachments. When fluctuations are large, the probability of attachment to all nodes becomes significant and is determined by the average degree of the network. More precisely, high fluctuations imply that new connections are established uniformly across all nodes.

For the formation of a $\Delta$-simplex, a newly added node must attach perfectly to a $(\Delta-1)$-simplex already present in the remainder of the graph. Such a connection must contain at least $\Delta$ links to the $(\Delta-1)$-simplex. There is therefore a combinatorial factor $\binom{m}{\Delta}$ corresponding to the number of different ways to form a $\Delta$-simplex from $m$ nodes. At time $t$, there are $t$ nodes available for attachment, and the average degree is defined as $\bar{k}_t \equiv \frac{k_t}{t}$, where $k_t$ is the total degree of the network up to time $t$. 

At the MF level, we assume that all nodes have the \textit{same} degree $\bar{k}_t$, so that the probability of a single attachment to a typical node is
\begin{equation}
p_{\text{attachment}}^{(0)} = \frac{\bar{k}_t}{k_t} = \frac{1}{t}.
\end{equation}
Note that $\bar{k}_t \propto t^{1/2}$ for the BA model. We must also take into account that, after a single attachment, the number of available nodes for further attachments is reduced. As a result, the average degree remains unchanged, while the total available degree changes according to $k_t^{(1)} = (t-1)\bar{k}_t$, so that
\begin{equation}
	p_{\text{attachment}}^{(1)}= \frac{\bar{k}_t}{k_t^{(1)}}=\frac{1}{t-1}.
\end{equation}
Upon $i$th attachment, this relation is generalized to:
\begin{equation}
p_{\text{attachemnt}}^{(i)}= \frac{\bar{k}_t}{k_t^{(i)}}=\frac{1}{t-i}.
\end{equation}
The probability increment $\delta P$ of successive $\Delta$ attachments to a $(\Delta-1)$-simplex is then given by multiplication of single probabilities, and also the number of $(\Delta-1)$ simplexes at time $t-1$:  
\begin{equation}
\begin{split}
 \delta P^{\text{MF}2}_\Delta(t) &= \binom{m}{\Delta}\cdot\left[\prod_{i=0}^{\Delta-1}(\Delta-i)p_{\text{attachemnt}}^{(i)}\right]\times P^{\text{MF}2}_\Delta(t-1)\\
&=\frac{m!}{(m-\Delta)!}\frac{(t-(\Delta-2))!}{t!}\times P^{\text{MF}2}_\Delta(t-1).
\end{split}
\end{equation}
The multiplicative $(\Delta-i)$ for $(i+1)$th attachment is the number of ways for the attachment at that stage. Using the Stirling's approximation, 
\begin{equation}
\ln t!\approx t\ln t-t,
\end{equation}
we find
\begin{equation}
\begin{split}
&\frac{(t-(\Delta-2))!}{t!}=\exp\left[\ln(t-(\Delta-2))!-\ln t! \right]\\
&\ \ \ =\exp\left[t\ln \frac{t-\Delta+2}{t}+(2-\Delta)\ln\left[t-\Delta+2\right]+\Delta-2\right]\\
&\ \ \ \approx \exp\left[\ln\left(t-\Delta+2\right)^{-(\Delta-2)}\right]\\
&\ \ \ \approx (t-t_0^{\text{MF}})^{-(\Delta-2)}\approx t^{-(\Delta-2)},
\end{split}
\end{equation}
where $t_0^{\text{MF}}\equiv \Delta-2$, and in the last step we used that fact that $t\gg t_0^{\text{MF}}$. This results in
\begin{equation}
	\delta P^{\text{MF}2}_\Delta(t) =\frac{m!}{(m-\Delta)!}t^{-(\Delta-2)}	P^{\text{MF}2}_\Delta(t-1).
\end{equation}
This is the leading term, while the sub-leading terms arise from the sub-leading terms obtained from the Stirling's approximation. We now use the fact that $ P^{\text{MF}2}_\Delta(t)\propto \Sigma_{\Delta}(t)$, and $\delta P^{\text{MF}2}_\Delta(t)\propto \delta\Sigma_\Delta(t)$, showing that ($\Sigma^{\text{MF}2}_\Delta(t-1)\approx\Sigma^{\text{MF}2}_\Delta(t)$)
\begin{equation}
	\frac{\delta \Sigma^{\text{MF}2}_\Delta(t)}{\Sigma^{\text{MF}2}_\Delta(t)} =\frac{m!}{(m-\Delta)!}t^{-(\Delta-2)}.
	\label{Eq:SigmaMF}
\end{equation}
To find the solutions for the above equation, we use the trial solution 
\begin{equation}
\delta \Sigma^{\text{MF}2}_\Delta(t)\propto t^{-\psi(\Delta)},
\end{equation}
one finds
\begin{equation}
	\Sigma^{\text{MF}2}_\Delta(t)=\sum_{s=1}^t\delta\Sigma^{\text{MF}2}_\Delta(s) \propto H(t,\psi),
\end{equation}
where $H(t,\psi)$ is the harmonic number given in Eq.~\ref{Eq:HarmonicNumber}. This tells us that
\begin{equation}
	\left. \Sigma^{\text{MF}2}_\Delta(t)\right|_{t\to \infty}\to \zeta(\psi)+\frac{t^{1-\psi}}{1-\psi}+O\left(t^{-\psi}\right),
\end{equation}
showing that $\Sigma^{\text{MF}2}_\Delta(t)$ saturates at $\zeta(\psi)$ at large enough times. Equation~\ref{Eq:SigmaMF} then tells us that
\begin{equation}
\left. \delta \Sigma^{\text{MF}2}_\Delta(t)\right|_{t\to\infty} =\zeta\left(\psi_{\text{MF}2}(\Delta)\right)\frac{m!}{(m-\Delta)!}t^{-\psi_{\text{MF}2}(\Delta)},
\end{equation}
where 
\begin{equation}
\psi_{\text{MF}2}(\Delta)=\Delta-2.
\end{equation}
For the case $m\gg\Delta$, using the Stirling's approximation we find  
\begin{equation}
\left. \delta \Sigma^{\text{MF}2}_\Delta(t)\right|_{t\to\infty,m\gg\Delta} =\zeta\left(\psi_{\text{MF}2}(\Delta)\right)(m-\Delta)^{-\Delta}t^{-\psi_{\text{MF}2}(\Delta)},
\end{equation}
which is valid for $m>\Delta$. Our numerical evaluations show that this approximation works better to large $\Delta$ values, see the main text.

\section{Fitting Details}
\begin{figure*}\label{fig:delta_sigma_fit_details}
    \includegraphics{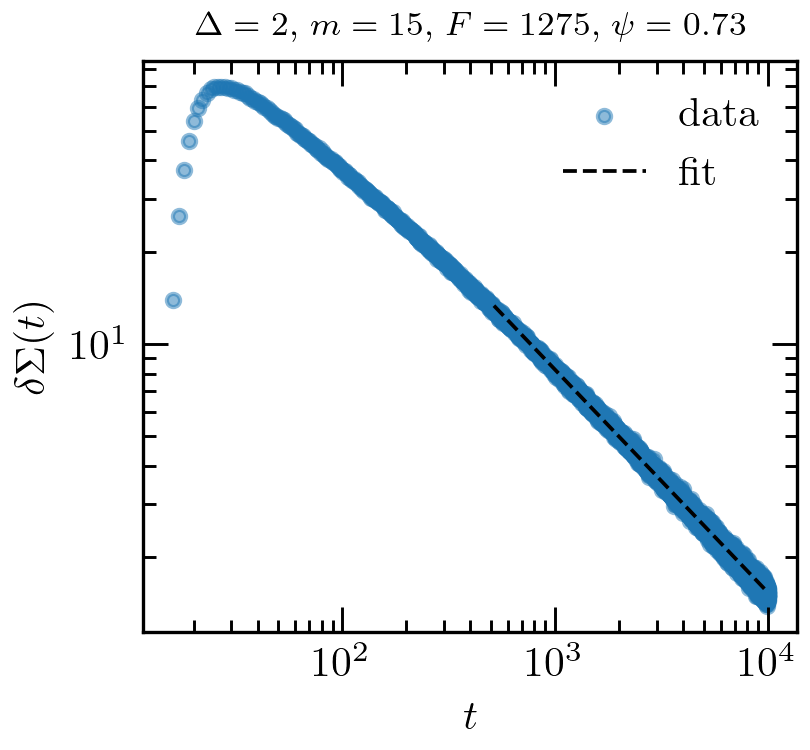}
    \includegraphics{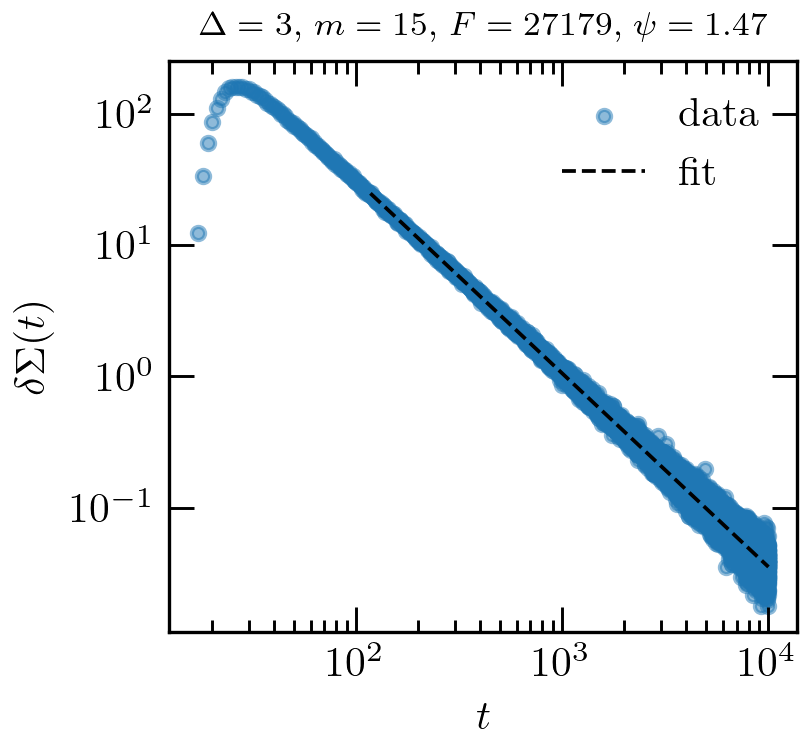}
    \includegraphics{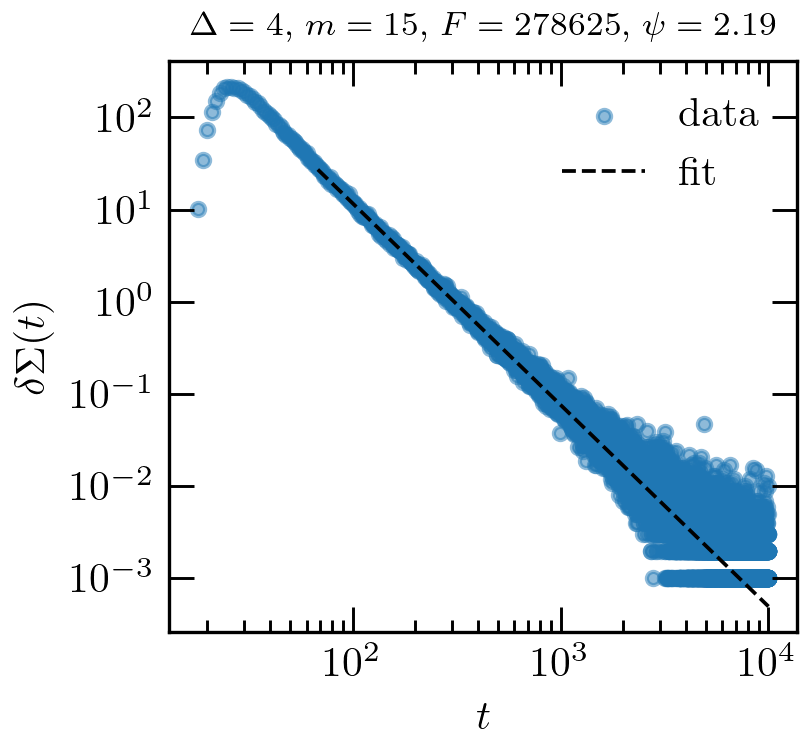}
    \includegraphics{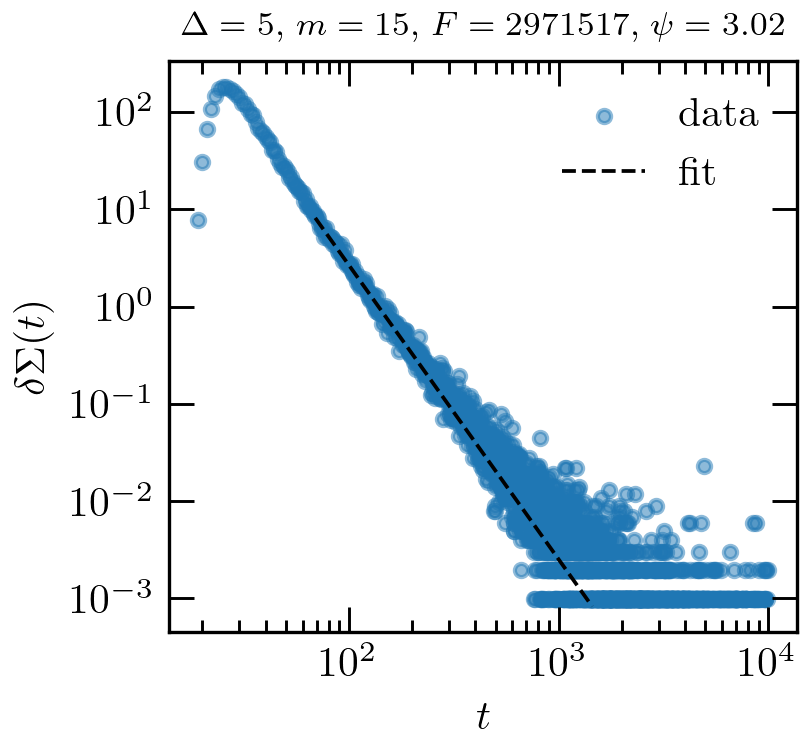}
	\caption{Fitting results for the number of simplexes, $\delta\Sigma_\Delta$, formed at each time step versus time for different simplex dimensions, $\Delta$. The connectivity parameter is fixed at $m=15$. The blue data points is the original data while the black dashed line is the fit. The model used is $\delta\Sigma_\Delta^{PL} = F\cdot t^{-\psi}$ and the values of the fitting parameters $F$ and $\psi$ are shown in the title, together with the value of $\Delta$.}
	\label{fig:delta_sigma_fit}
\end{figure*}

\begin{figure*}\label{fig:F_sigma_comparison}
    \includegraphics{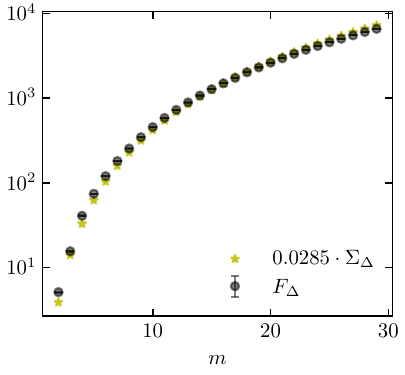}
    \includegraphics{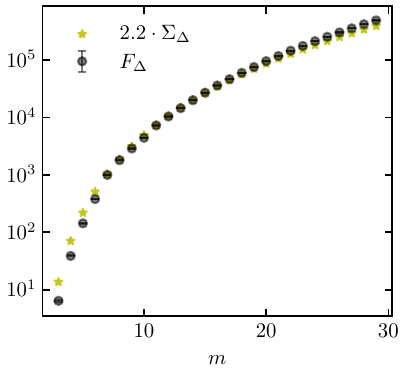}
    \includegraphics{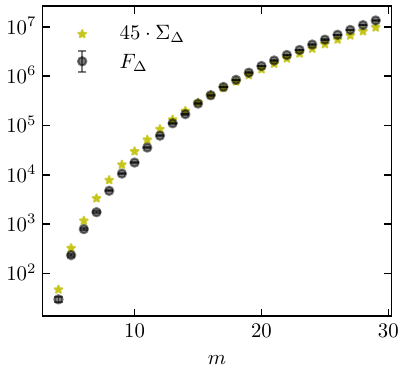}
    \includegraphics{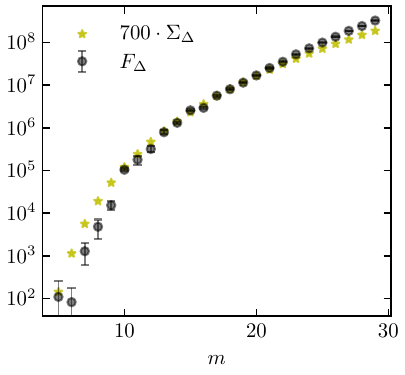}
	\caption{Comparison of the values of $F(m)$ and $\Sigma_\Delta(m)$ for different values of $\Delta$. As described in the text, we expect that $F(m)\propto\Sigma_\Delta^{PL}(m)$ and the question is whether it is also true that $F(m)\propto\Sigma_\Delta(m)$. From the resutls in this figure we can conclude that $\Sigma_\Delta^{PL}(m)$ is a good approximation for $\Sigma_\Delta(m)$. We rescaled the values of $\Sigma$ so that the two data sets come as close as possible. The values of the proportionality constant, the factor that we used to rescale the values of $F$ is shown within the each graph. }
	\label{fig:dF_fit}
\end{figure*}

The details of the fit for the parameter $\delta\Sigma$ in several dimensions are provided here. These include examples of fits for the parameters $F$ and $\Sigma$, introduced in Eqs.~\eqref{eqt:sigma_delta_t_m} and~\eqref{eq:total_simplexes} in the main text, together with a comparison between them. The results are shown in Figs.~\ref{fig:delta_sigma_fit} and~\ref{fig:dF_fit}.

\section{Complex Networks and Topological Concepts}\label{SEC:tda}
Here, we overview of the main mathematical concepts related to the topology of complex networks.

\subsection{Simplicial Complexes in Networks}
To analyse the topology of complex networks, one can first approximate it with a simplicial complex,  a combinatorial representation that preserves the topological essence of the network built from simplexes (points, edges, triangles, and higher-dimensional analogues). In this context: 0-simplexes are the nodes (vertices) of the network. 1-simplexes are the edges between nodes.
Higher-dimensional simplexes (such as triangles, or 2-simplexes, and tetrahedra, or 3-simplexes) are related to higher order connections involving more than two nodes. The following definitions are crucial in our discussions: 
\begin{itemize}
    \item The $n$-simplex denoted by $\sigma_n=(v_0\: v_1\: ...\: v_n)$ in a space with dimension $\mathbb{R}^n$ is formally defined as
\begin{equation}
	\sigma_n = \{x\in \mathbb{R}^n \, | \, x=\sum^{n}_{i=0}\lambda_iv_i, \, \lambda_i\geq0, \, \sum^{n}_{i=0}\lambda_i =1\},
\end{equation}
where $\left\lbrace v_i\right\rbrace_{i=1}^n $ represents the set of nodes, and must be geometrically independent in order to represent an $n$-dimensional object. The $\lambda_i$s are some coefficients that satisfy the conditions of the above equation.

\item A \textit{$p$-face} of a $n$-simplex is a simplex itself made by $p+1$ points and it is a subset of the $n$-simplex such that $0\leq p\leq n$. 

\item A simplicial complex $K$ is a set whose elements are the simplexes with different dimensions. The dimension of a simplicial complex is defined to be the dimension of its largest simplex.

\end{itemize}
 
 Two conditions are necessary in order to make a simplicial complex which is going to be homeomorphism to the underlying topological object;
\begin{itemize}
	\item Every face of each simplex of a simplicial complex $K$ also belongs to $K$.
	\item The intersection of any pair of two simplexes in $K$ is either empty or a face of those two simplexes.
\end{itemize}

\subsection{Homology Groups and Betti Numbers}
Homology provides a way to quantify the topological features of a network at different dimensions. A $d$-dimensional homology focuses on $d$-dimensional voids; when $d=0$ and $d=1$ concern the number of connected components and the cycles or loops in the network respectively, while higher-dimensional homology describes voids in higher-order simplicial complexes. In this regard, the \textit{Betti numbers} ($\beta_d$) quantify the number of independent topological features: The 0th Betti number ($\beta_0$) counts the number of connected components, while $\beta_1$ counts the number of independent loops.\\

In order to obtain the homology groups of a simplicial complex that is homeomorphic to a graph, one has to understand what \textit{chain groups}, \textit{cycle groups} and \textit{boundry groups} of a simplicial complex are. Let $K$ be our $n$-dimensional simplicial complex. The $d$-chain group $C_d(K)$, $0\leq d\leq n$, with the addition operation is the set of all $d$-chains denoted by $c$ and is represented as 
\begin{equation}
	C_d(K) = \lbrace c \: \vert \: c=\sum^{N_d}_{i=1}c_i\sigma^i_{d}, \: c_i\in \mathbb{Z} \rbrace
\end{equation}
where $N_d$ is the total number of $d$-simplexes available in the simplicial complex $K$, and $\sigma_d^i$ is the $i$th $d$-dimensional simplex in $K$.
 The unit element of the $d$-chain group $C_d(K)$ is $0 = \sum_i 0.\sigma^i_{d}$. Also, each element $c$ has an inverse element of $-c = \sum^{N_d}_{i=1}(-c_i)\sigma^i_{d}$. The $C_d(K)$ has the properties of free Abelian groups, and its rank is $N_d$.

The $d$-cycle group denoted by $Z_d(K)$ is defined as 
\begin{equation}
	Z_d(K)=\lbrace c \: \vert \: \partial_dc=0, \: c\in C_d(K)\rbrace
	\label{eqt_r_cycle}
\end{equation}
where $\partial_d$ is called the boundary operater and it acts on simplexes as
\begin{equation}
	\partial_d\sigma_d=\sum^d_{i=0}(-1)^i(v_0\: v_1\: ... \: \hat{v}_i\: ...\: v_n)
	\label{eqt_bndr_operator}
\end{equation}
where the point $\hat{v}_i$ is ignored. The $d$-cycle group is a subgroup of $d$-chain group. The elements of the $d$-cycle group are called $d$-cycles. By combining the Eqs. (\ref{eqt_r_cycle}) and (\ref{eqt_bndr_operator}) one can find that the boundary operator $\partial_d$ is actually a map from $C_d(K)$ to $C_{d-1}(K)$
\begin{equation}
	\partial_d : C_d(K) \to C_{d-1}(K).
\end{equation}
The $d$-boundary group is defined as 
\begin{equation}
	B_d(K)= \lbrace c\: \vert \: c=\partial_{d+1}c', \: c'\in C_{d+1}(K)\rbrace.
\end{equation}
For an $n$-dimensional simplicial complex $K$ the $d$th homology group $H_d(K)$, $0 \leq d \leq n$, associated with $K$ is defined by
\begin{equation}
	H_d(K)\equiv Z_d(K)/B_d(K).
\end{equation}
which is topological invariant. For a simplicial complex $K$, the $d$th Betti number, $\beta_d(K)$, is defined by
\begin{equation}
	\beta_d(K)\equiv \mathrm{rank}\, H_d(K).
\end{equation}

\bibliography{refs}

\begin{thebibliography}{51}%
\makeatletter
\providecommand \@ifxundefined [1]{%
 \@ifx{#1\undefined}
}%
\providecommand \@ifnum [1]{%
 \ifnum #1\expandafter \@firstoftwo
 \else \expandafter \@secondoftwo
 \fi
}%
\providecommand \@ifx [1]{%
 \ifx #1\expandafter \@firstoftwo
 \else \expandafter \@secondoftwo
 \fi
}%
\providecommand \natexlab [1]{#1}%
\providecommand \enquote  [1]{``#1''}%
\providecommand \bibnamefont  [1]{#1}%
\providecommand \bibfnamefont [1]{#1}%
\providecommand \citenamefont [1]{#1}%
\providecommand \href@noop [0]{\@secondoftwo}%
\providecommand \href [0]{\begingroup \@sanitize@url \@href}%
\providecommand \@href[1]{\@@startlink{#1}\@@href}%
\providecommand \@@href[1]{\endgroup#1\@@endlink}%
\providecommand \@sanitize@url [0]{\catcode `\\12\catcode `\$12\catcode
  `\&12\catcode `\#12\catcode `\^12\catcode `\_12\catcode `\%12\relax}%
\providecommand \@@startlink[1]{}%
\providecommand \@@endlink[0]{}%
\providecommand \url  [0]{\begingroup\@sanitize@url \@url }%
\providecommand \@url [1]{\endgroup\@href {#1}{\urlprefix }}%
\providecommand \urlprefix  [0]{URL }%
\providecommand \Eprint [0]{\href }%
\providecommand \doibase [0]{https://doi.org/}%
\providecommand \selectlanguage [0]{\@gobble}%
\providecommand \bibinfo  [0]{\@secondoftwo}%
\providecommand \bibfield  [0]{\@secondoftwo}%
\providecommand \translation [1]{[#1]}%
\providecommand \BibitemOpen [0]{}%
\providecommand \bibitemStop [0]{}%
\providecommand \bibitemNoStop [0]{.\EOS\space}%
\providecommand \EOS [0]{\spacefactor3000\relax}%
\providecommand \BibitemShut  [1]{\csname bibitem#1\endcsname}%
\let\auto@bib@innerbib\@empty
\bibitem [{\citenamefont {Albert}\ and\ \citenamefont
  {Barab\'asi}(2002)}]{albert2002statistical}%
  \BibitemOpen
  \bibfield  {author} {\bibinfo {author} {\bibfnamefont {R.}~\bibnamefont
  {Albert}}\ and\ \bibinfo {author} {\bibfnamefont {A.-L.}\ \bibnamefont
  {Barab\'asi}},\ }\bibfield  {title} {\bibinfo {title} {Statistical mechanics
  of complex networks},\ }\href {https://doi.org/10.1103/RevModPhys.74.47}
  {\bibfield  {journal} {\bibinfo  {journal} {Rev. Mod. Phys.}\ }\textbf
  {\bibinfo {volume} {74}},\ \bibinfo {pages} {47} (\bibinfo {year}
  {2002})}\BibitemShut {NoStop}%
\bibitem [{\citenamefont {Boccaletti}\ \emph {et~al.}(2006)\citenamefont
  {Boccaletti}, \citenamefont {Latora}, \citenamefont {Moreno}, \citenamefont
  {Chavez},\ and\ \citenamefont {Hwang}}]{BOCCALETTI2006175}%
  \BibitemOpen
  \bibfield  {author} {\bibinfo {author} {\bibfnamefont {S.}~\bibnamefont
  {Boccaletti}}, \bibinfo {author} {\bibfnamefont {V.}~\bibnamefont {Latora}},
  \bibinfo {author} {\bibfnamefont {Y.}~\bibnamefont {Moreno}}, \bibinfo
  {author} {\bibfnamefont {M.}~\bibnamefont {Chavez}},\ and\ \bibinfo {author}
  {\bibfnamefont {D.-U.}\ \bibnamefont {Hwang}},\ }\bibfield  {title} {\bibinfo
  {title} {Complex networks: Structure and dynamics},\ }\href
  {https://doi.org/https://doi.org/10.1016/j.physrep.2005.10.009} {\bibfield
  {journal} {\bibinfo  {journal} {Physics Reports}\ }\textbf {\bibinfo {volume}
  {424}},\ \bibinfo {pages} {175} (\bibinfo {year} {2006})}\BibitemShut
  {NoStop}%
\bibitem [{\citenamefont {Newman}(2003)}]{NewmanS003614450342480}%
  \BibitemOpen
  \bibfield  {author} {\bibinfo {author} {\bibfnamefont {M.~E.~J.}\
  \bibnamefont {Newman}},\ }\bibfield  {title} {\bibinfo {title} {The structure
  and function of complex networks},\ }\href
  {https://doi.org/10.1137/S003614450342480} {\bibfield  {journal} {\bibinfo
  {journal} {SIAM Review}\ }\textbf {\bibinfo {volume} {45}},\ \bibinfo {pages}
  {167} (\bibinfo {year} {2003})}\BibitemShut {NoStop}%
\bibitem [{\citenamefont {Barabási}\ and\ \citenamefont
  {Albert}(1999)}]{barabasi1999emergence}%
  \BibitemOpen
  \bibfield  {author} {\bibinfo {author} {\bibfnamefont {A.-L.}\ \bibnamefont
  {Barabási}}\ and\ \bibinfo {author} {\bibfnamefont {R.}~\bibnamefont
  {Albert}},\ }\bibfield  {title} {\bibinfo {title} {Emergence of scaling in
  random networks},\ }\href {https://doi.org/10.1126/science.286.5439.509}
  {\bibfield  {journal} {\bibinfo  {journal} {Science}\ }\textbf {\bibinfo
  {volume} {286}},\ \bibinfo {pages} {509} (\bibinfo {year}
  {1999})}\BibitemShut {NoStop}%
\bibitem [{\citenamefont {Pastor-Satorras}\ \emph {et~al.}(2001)\citenamefont
  {Pastor-Satorras}, \citenamefont {V\'azquez},\ and\ \citenamefont
  {Vespignani}}]{pastor2001dynamical}%
  \BibitemOpen
  \bibfield  {author} {\bibinfo {author} {\bibfnamefont {R.}~\bibnamefont
  {Pastor-Satorras}}, \bibinfo {author} {\bibfnamefont {A.}~\bibnamefont
  {V\'azquez}},\ and\ \bibinfo {author} {\bibfnamefont {A.}~\bibnamefont
  {Vespignani}},\ }\bibfield  {title} {\bibinfo {title} {Dynamical and
  correlation properties of the internet},\ }\href
  {https://doi.org/10.1103/PhysRevLett.87.258701} {\bibfield  {journal}
  {\bibinfo  {journal} {Phys. Rev. Lett.}\ }\textbf {\bibinfo {volume} {87}},\
  \bibinfo {pages} {258701} (\bibinfo {year} {2001})}\BibitemShut {NoStop}%
\bibitem [{\citenamefont {Fotouhi}\ and\ \citenamefont
  {Rabbat}(2013)}]{fotouhi2013degree}%
  \BibitemOpen
  \bibfield  {author} {\bibinfo {author} {\bibfnamefont {B.}~\bibnamefont
  {Fotouhi}}\ and\ \bibinfo {author} {\bibfnamefont {M.~G.}\ \bibnamefont
  {Rabbat}},\ }\bibfield  {title} {\bibinfo {title} {Degree correlation in
  scale-free graphs},\ }\href {https://doi.org/10.1140/epjb/e2013-40920-6}
  {\bibfield  {journal} {\bibinfo  {journal} {The European Physical Journal B}\
  }\textbf {\bibinfo {volume} {86}},\ \bibinfo {pages} {510} (\bibinfo {year}
  {2013})}\BibitemShut {NoStop}%
\bibitem [{\citenamefont {Szab\'o}\ \emph {et~al.}(2003)\citenamefont
  {Szab\'o}, \citenamefont {Alava},\ and\ \citenamefont
  {Kert\'esz}}]{szabo2003structural}%
  \BibitemOpen
  \bibfield  {author} {\bibinfo {author} {\bibfnamefont {G.}~\bibnamefont
  {Szab\'o}}, \bibinfo {author} {\bibfnamefont {M.}~\bibnamefont {Alava}},\
  and\ \bibinfo {author} {\bibfnamefont {J.}~\bibnamefont {Kert\'esz}},\
  }\bibfield  {title} {\bibinfo {title} {Structural transitions in scale-free
  networks},\ }\href {https://doi.org/10.1103/PhysRevE.67.056102} {\bibfield
  {journal} {\bibinfo  {journal} {Phys. Rev. E}\ }\textbf {\bibinfo {volume}
  {67}},\ \bibinfo {pages} {056102} (\bibinfo {year} {2003})}\BibitemShut
  {NoStop}%
\bibitem [{\citenamefont {Krapivsky}\ \emph {et~al.}(2000)\citenamefont
  {Krapivsky}, \citenamefont {Redner},\ and\ \citenamefont
  {Leyvraz}}]{krapivsky2000connectivity}%
  \BibitemOpen
  \bibfield  {author} {\bibinfo {author} {\bibfnamefont {P.~L.}\ \bibnamefont
  {Krapivsky}}, \bibinfo {author} {\bibfnamefont {S.}~\bibnamefont {Redner}},\
  and\ \bibinfo {author} {\bibfnamefont {F.}~\bibnamefont {Leyvraz}},\
  }\bibfield  {title} {\bibinfo {title} {Connectivity of growing random
  networks},\ }\href {https://doi.org/10.1103/PhysRevLett.85.4629} {\bibfield
  {journal} {\bibinfo  {journal} {Phys. Rev. Lett.}\ }\textbf {\bibinfo
  {volume} {85}},\ \bibinfo {pages} {4629} (\bibinfo {year}
  {2000})}\BibitemShut {NoStop}%
\bibitem [{\citenamefont {Horak}\ \emph {et~al.}(2009)\citenamefont {Horak},
  \citenamefont {Maleti{\'c}},\ and\ \citenamefont
  {Rajkovi{\'c}}}]{horak2009persistent}%
  \BibitemOpen
  \bibfield  {author} {\bibinfo {author} {\bibfnamefont {D.}~\bibnamefont
  {Horak}}, \bibinfo {author} {\bibfnamefont {S.}~\bibnamefont {Maleti{\'c}}},\
  and\ \bibinfo {author} {\bibfnamefont {M.}~\bibnamefont {Rajkovi{\'c}}},\
  }\bibfield  {title} {\bibinfo {title} {Persistent homology of complex
  networks},\ }\href {https://doi.org/10.1088/1742-5468/2009/03/P03034}
  {\bibfield  {journal} {\bibinfo  {journal} {Journal of Statistical Mechanics:
  Theory and Experiment}\ }\textbf {\bibinfo {volume} {2009}},\ \bibinfo
  {pages} {P03034} (\bibinfo {year} {2009})}\BibitemShut {NoStop}%
\bibitem [{\citenamefont {Sizemore}\ \emph {et~al.}(2018)\citenamefont
  {Sizemore}, \citenamefont {Giusti}, \citenamefont {Kahn}, \citenamefont
  {Vettel}, \citenamefont {Betzel},\ and\ \citenamefont
  {Bassett}}]{sizemore2018cliques}%
  \BibitemOpen
  \bibfield  {author} {\bibinfo {author} {\bibfnamefont {A.~E.}\ \bibnamefont
  {Sizemore}}, \bibinfo {author} {\bibfnamefont {C.}~\bibnamefont {Giusti}},
  \bibinfo {author} {\bibfnamefont {A.}~\bibnamefont {Kahn}}, \bibinfo {author}
  {\bibfnamefont {J.~M.}\ \bibnamefont {Vettel}}, \bibinfo {author}
  {\bibfnamefont {R.~F.}\ \bibnamefont {Betzel}},\ and\ \bibinfo {author}
  {\bibfnamefont {D.~S.}\ \bibnamefont {Bassett}},\ }\bibfield  {title}
  {\bibinfo {title} {Cliques and cavities in the human connectome},\ }\href
  {https://doi.org/10.1007/s10827-017-0672-6} {\bibfield  {journal} {\bibinfo
  {journal} {Journal of computational neuroscience}\ }\textbf {\bibinfo
  {volume} {44}},\ \bibinfo {pages} {115} (\bibinfo {year} {2018})}\BibitemShut
  {NoStop}%
\bibitem [{\citenamefont {Mill{\'a}n}\ \emph {et~al.}(2020)\citenamefont
  {Mill{\'a}n}, \citenamefont {Torres},\ and\ \citenamefont
  {Bianconi}}]{millan2020explosive}%
  \BibitemOpen
  \bibfield  {author} {\bibinfo {author} {\bibfnamefont {A.~P.}\ \bibnamefont
  {Mill{\'a}n}}, \bibinfo {author} {\bibfnamefont {J.~J.}\ \bibnamefont
  {Torres}},\ and\ \bibinfo {author} {\bibfnamefont {G.}~\bibnamefont
  {Bianconi}},\ }\bibfield  {title} {\bibinfo {title} {Explosive higher-order
  kuramoto dynamics on simplicial complexes},\ }\href
  {https://doi.org/10.1103/PhysRevLett.124.218301} {\bibfield  {journal}
  {\bibinfo  {journal} {Physical Review Letters}\ }\textbf {\bibinfo {volume}
  {124}},\ \bibinfo {pages} {218301} (\bibinfo {year} {2020})}\BibitemShut
  {NoStop}%
\bibitem [{\citenamefont {Iacopini}\ \emph {et~al.}(2019)\citenamefont
  {Iacopini}, \citenamefont {Petri}, \citenamefont {Barrat},\ and\
  \citenamefont {Latora}}]{iacopini2019simplicial}%
  \BibitemOpen
  \bibfield  {author} {\bibinfo {author} {\bibfnamefont {I.}~\bibnamefont
  {Iacopini}}, \bibinfo {author} {\bibfnamefont {G.}~\bibnamefont {Petri}},
  \bibinfo {author} {\bibfnamefont {A.}~\bibnamefont {Barrat}},\ and\ \bibinfo
  {author} {\bibfnamefont {V.}~\bibnamefont {Latora}},\ }\bibfield  {title}
  {\bibinfo {title} {Simplicial models of social contagion},\ }\href
  {https://doi.org/10.1038/s41467-019-10431-6} {\bibfield  {journal} {\bibinfo
  {journal} {Nature communications}\ }\textbf {\bibinfo {volume} {10}},\
  \bibinfo {pages} {2485} (\bibinfo {year} {2019})}\BibitemShut {NoStop}%
\bibitem [{\citenamefont {Giusti}\ \emph
  {et~al.}(2016{\natexlab{a}})\citenamefont {Giusti}, \citenamefont {Ghrist},\
  and\ \citenamefont {Bassett}}]{giusti2016two}%
  \BibitemOpen
  \bibfield  {author} {\bibinfo {author} {\bibfnamefont {C.}~\bibnamefont
  {Giusti}}, \bibinfo {author} {\bibfnamefont {R.}~\bibnamefont {Ghrist}},\
  and\ \bibinfo {author} {\bibfnamefont {D.~S.}\ \bibnamefont {Bassett}},\
  }\bibfield  {title} {\bibinfo {title} {Two’s company, three (or more) is a
  simplex: Algebraic-topological tools for understanding higher-order structure
  in neural data},\ }\href {https://doi.org/10.1007/s10827-016-0608-6}
  {\bibfield  {journal} {\bibinfo  {journal} {Journal of computational
  neuroscience}\ }\textbf {\bibinfo {volume} {41}},\ \bibinfo {pages} {1}
  (\bibinfo {year} {2016}{\natexlab{a}})}\BibitemShut {NoStop}%
\bibitem [{\citenamefont {Zhao}\ \emph {et~al.}(2022)\citenamefont {Zhao},
  \citenamefont {Li}, \citenamefont {Peng}, \citenamefont {Zhong},\ and\
  \citenamefont {Wang}}]{zhao2022higher}%
  \BibitemOpen
  \bibfield  {author} {\bibinfo {author} {\bibfnamefont {D.}~\bibnamefont
  {Zhao}}, \bibinfo {author} {\bibfnamefont {R.}~\bibnamefont {Li}}, \bibinfo
  {author} {\bibfnamefont {H.}~\bibnamefont {Peng}}, \bibinfo {author}
  {\bibfnamefont {M.}~\bibnamefont {Zhong}},\ and\ \bibinfo {author}
  {\bibfnamefont {W.}~\bibnamefont {Wang}},\ }\bibfield  {title} {\bibinfo
  {title} {Higher-order percolation in simplicial complexes},\ }\href
  {https://doi.org/10.1016/j.chaos.2021.111701} {\bibfield  {journal} {\bibinfo
   {journal} {Chaos, Solitons \& Fractals}\ }\textbf {\bibinfo {volume}
  {155}},\ \bibinfo {pages} {111701} (\bibinfo {year} {2022})}\BibitemShut
  {NoStop}%
\bibitem [{\citenamefont {Edelsbrunner}\ \emph {et~al.}(2002)\citenamefont
  {Edelsbrunner}, \citenamefont {Letscher},\ and\ \citenamefont
  {Zomorodian}}]{edelsbrunner2002topological}%
  \BibitemOpen
  \bibfield  {author} {\bibinfo {author} {\bibnamefont {Edelsbrunner}},
  \bibinfo {author} {\bibnamefont {Letscher}},\ and\ \bibinfo {author}
  {\bibnamefont {Zomorodian}},\ }\bibfield  {title} {\bibinfo {title}
  {Topological persistence and simplification},\ }\href
  {https://doi.org/10.1007/s00454-002-2885-2} {\bibfield  {journal} {\bibinfo
  {journal} {Discrete \& computational geometry}\ }\textbf {\bibinfo {volume}
  {28}},\ \bibinfo {pages} {511} (\bibinfo {year} {2002})}\BibitemShut
  {NoStop}%
\bibitem [{\citenamefont {Edelsbrunner}\ and\ \citenamefont
  {Harer}(2010)}]{edelsbrunner2010computational}%
  \BibitemOpen
  \bibfield  {author} {\bibinfo {author} {\bibfnamefont {H.}~\bibnamefont
  {Edelsbrunner}}\ and\ \bibinfo {author} {\bibfnamefont {J.}~\bibnamefont
  {Harer}},\ }\href {https://doi.org/10.1090/mbk/069} {\emph {\bibinfo {title}
  {Computational topology: an introduction}}}\ (\bibinfo  {publisher} {American
  Mathematical Soc.},\ \bibinfo {year} {2010})\BibitemShut {NoStop}%
\bibitem [{\citenamefont {Shi}\ \emph {et~al.}(2021)\citenamefont {Shi},
  \citenamefont {Chen}, \citenamefont {Sun}, \citenamefont {Chen},
  \citenamefont {Ma}, \citenamefont {Lou},\ and\ \citenamefont
  {Chen}}]{shi2021}%
  \BibitemOpen
  \bibfield  {author} {\bibinfo {author} {\bibfnamefont {D.}~\bibnamefont
  {Shi}}, \bibinfo {author} {\bibfnamefont {Z.}~\bibnamefont {Chen}}, \bibinfo
  {author} {\bibfnamefont {X.}~\bibnamefont {Sun}}, \bibinfo {author}
  {\bibfnamefont {Q.}~\bibnamefont {Chen}}, \bibinfo {author} {\bibfnamefont
  {C.}~\bibnamefont {Ma}}, \bibinfo {author} {\bibfnamefont {Y.}~\bibnamefont
  {Lou}},\ and\ \bibinfo {author} {\bibfnamefont {G.}~\bibnamefont {Chen}},\
  }\bibfield  {title} {\bibinfo {title} {Computing cliques and cavities in
  networks},\ }\href {https://doi.org/10.1038/s42005-021-00748-4} {\bibfield
  {journal} {\bibinfo  {journal} {Communications Physics}\ }\textbf {\bibinfo
  {volume} {4}},\ \bibinfo {pages} {249} (\bibinfo {year} {2021})}\BibitemShut
  {NoStop}%
\bibitem [{\citenamefont {Reimann}\ \emph {et~al.}(2017)\citenamefont
  {Reimann}, \citenamefont {Nolte}, \citenamefont {Scolamiero}, \citenamefont
  {Turner}, \citenamefont {Perin}, \citenamefont {G.}, \citenamefont {Dłotko},
  \citenamefont {Levi},\ and\ \citenamefont {H}}]{reimann2017}%
  \BibitemOpen
  \bibfield  {author} {\bibinfo {author} {\bibfnamefont {M.}~\bibnamefont
  {Reimann}}, \bibinfo {author} {\bibfnamefont {M.}~\bibnamefont {Nolte}},
  \bibinfo {author} {\bibfnamefont {M.}~\bibnamefont {Scolamiero}}, \bibinfo
  {author} {\bibfnamefont {K.}~\bibnamefont {Turner}}, \bibinfo {author}
  {\bibfnamefont {R.}~\bibnamefont {Perin}}, \bibinfo {author} {\bibfnamefont
  {C.}~\bibnamefont {G.}}, \bibinfo {author} {\bibfnamefont {P.}~\bibnamefont
  {Dłotko}}, \bibinfo {author} {\bibfnamefont {K.}~\bibnamefont {Levi},
  \bibfnamefont {R.~Hess}},\ and\ \bibinfo {author} {\bibfnamefont
  {M.}~\bibnamefont {H}},\ }\bibfield  {title} {\bibinfo {title} {Cliques of
  neurons bound into cavities provide a missing link between structure and
  function},\ }\href {https://doi.org/10.3389/fncom.2017.00048} {\bibfield
  {journal} {\bibinfo  {journal} {Front. Comput. Neurosci.}\ }\textbf {\bibinfo
  {volume} {11:48}} (\bibinfo {year} {2017})}\BibitemShut {NoStop}%
\bibitem [{\citenamefont {Giusti}\ \emph
  {et~al.}(2016{\natexlab{b}})\citenamefont {Giusti}, \citenamefont {Ghrist},\
  and\ \citenamefont {Bassett}}]{giusti2016}%
  \BibitemOpen
  \bibfield  {author} {\bibinfo {author} {\bibfnamefont {C.}~\bibnamefont
  {Giusti}}, \bibinfo {author} {\bibfnamefont {R.}~\bibnamefont {Ghrist}},\
  and\ \bibinfo {author} {\bibfnamefont {D.~S.}\ \bibnamefont {Bassett}},\
  }\bibfield  {title} {\bibinfo {title} {Two's company, three (or more) is a
  simplex},\ }\href {https://doi.org/10.1007/s10827-016-0608-6} {\bibfield
  {journal} {\bibinfo  {journal} {Journal of Computational Neuroscience}\
  }\textbf {\bibinfo {volume} {41}},\ \bibinfo {pages} {1} (\bibinfo {year}
  {2016}{\natexlab{b}})}\BibitemShut {NoStop}%
\bibitem [{\citenamefont {Gameiro}\ \emph {et~al.}(2015)\citenamefont
  {Gameiro}, \citenamefont {Hiraoka}, \citenamefont {Izumi}, \citenamefont
  {Kramar}, \citenamefont {Mischaikow},\ and\ \citenamefont
  {Nanda}}]{gameiro2015}%
  \BibitemOpen
  \bibfield  {author} {\bibinfo {author} {\bibfnamefont {M.}~\bibnamefont
  {Gameiro}}, \bibinfo {author} {\bibfnamefont {Y.}~\bibnamefont {Hiraoka}},
  \bibinfo {author} {\bibfnamefont {S.}~\bibnamefont {Izumi}}, \bibinfo
  {author} {\bibfnamefont {M.}~\bibnamefont {Kramar}}, \bibinfo {author}
  {\bibfnamefont {K.}~\bibnamefont {Mischaikow}},\ and\ \bibinfo {author}
  {\bibfnamefont {V.}~\bibnamefont {Nanda}},\ }\bibfield  {title} {\bibinfo
  {title} {A topological measurement of protein compressibility},\ }\href
  {https://doi.org/10.1007/s13160-014-0153-5} {\bibfield  {journal} {\bibinfo
  {journal} {Japan Journal of Industrial and Applied Mathematics}\ }\textbf
  {\bibinfo {volume} {32}},\ \bibinfo {pages} {1} (\bibinfo {year}
  {2015})}\BibitemShut {NoStop}%
\bibitem [{\citenamefont {Jiang}\ \emph {et~al.}(2018)\citenamefont {Jiang},
  \citenamefont {Tsuji},\ and\ \citenamefont {Shirai}}]{jiang2018}%
  \BibitemOpen
  \bibfield  {author} {\bibinfo {author} {\bibfnamefont {F.}~\bibnamefont
  {Jiang}}, \bibinfo {author} {\bibfnamefont {T.}~\bibnamefont {Tsuji}},\ and\
  \bibinfo {author} {\bibfnamefont {T.}~\bibnamefont {Shirai}},\ }\bibfield
  {title} {\bibinfo {title} {Pore geometry characterization by persistent
  homology theory},\ }\href
  {https://doi.org/https://doi.org/10.1029/2017WR021864} {\bibfield  {journal}
  {\bibinfo  {journal} {Water Resources Research}\ }\textbf {\bibinfo {volume}
  {54}},\ \bibinfo {pages} {4150} (\bibinfo {year} {2018})}\BibitemShut
  {NoStop}%
\bibitem [{\citenamefont {Lee}\ \emph {et~al.}(2017)\citenamefont {Lee},
  \citenamefont {Barthel}, \citenamefont {D{\l}otko}, \citenamefont {Moosavi},
  \citenamefont {Hess},\ and\ \citenamefont {Smit}}]{yongjin2017}%
  \BibitemOpen
  \bibfield  {author} {\bibinfo {author} {\bibfnamefont {Y.}~\bibnamefont
  {Lee}}, \bibinfo {author} {\bibfnamefont {S.~D.}\ \bibnamefont {Barthel}},
  \bibinfo {author} {\bibfnamefont {P.}~\bibnamefont {D{\l}otko}}, \bibinfo
  {author} {\bibfnamefont {S.~M.}\ \bibnamefont {Moosavi}}, \bibinfo {author}
  {\bibfnamefont {K.}~\bibnamefont {Hess}},\ and\ \bibinfo {author}
  {\bibfnamefont {B.}~\bibnamefont {Smit}},\ }\bibfield  {title} {\bibinfo
  {title} {Quantifying similarity of pore-geometry in nanoporous materials},\
  }\href {https://doi.org/10.1038/ncomms15396} {\bibfield  {journal} {\bibinfo
  {journal} {Nature Communications}\ }\textbf {\bibinfo {volume} {8}},\
  \bibinfo {pages} {15396} (\bibinfo {year} {2017})}\BibitemShut {NoStop}%
\bibitem [{\citenamefont {Szymanski}\ \emph {et~al.}(2025)\citenamefont
  {Szymanski}, \citenamefont {Smith}, \citenamefont {Daoutidis},\ and\
  \citenamefont {Bartel}}]{szymanski2025}%
  \BibitemOpen
  \bibfield  {author} {\bibinfo {author} {\bibfnamefont {N.~J.}\ \bibnamefont
  {Szymanski}}, \bibinfo {author} {\bibfnamefont {A.}~\bibnamefont {Smith}},
  \bibinfo {author} {\bibfnamefont {P.}~\bibnamefont {Daoutidis}},\ and\
  \bibinfo {author} {\bibfnamefont {C.~J.}\ \bibnamefont {Bartel}},\ }\bibfield
   {title} {\bibinfo {title} {Topological descriptors for the electron density
  of inorganic solids},\ }\href
  {https://doi.org/10.1021/acsmaterialslett.5c00390} {\bibfield  {journal}
  {\bibinfo  {journal} {ACS Materials Letters}\ }\textbf {\bibinfo {volume}
  {7}},\ \bibinfo {pages} {2158} (\bibinfo {year} {2025})}\BibitemShut
  {NoStop}%
\bibitem [{\citenamefont {Serrano}\ and\ \citenamefont
  {G{\'o}mez}(2020)}]{serrano2020centrality}%
  \BibitemOpen
  \bibfield  {author} {\bibinfo {author} {\bibfnamefont {D.~H.}\ \bibnamefont
  {Serrano}}\ and\ \bibinfo {author} {\bibfnamefont {D.~S.}\ \bibnamefont
  {G{\'o}mez}},\ }\bibfield  {title} {\bibinfo {title} {Centrality measures in
  simplicial complexes: Applications of topological data analysis to network
  science},\ }\href {https://doi.org/10.1016/j.amc.2020.125331} {\bibfield
  {journal} {\bibinfo  {journal} {Applied Mathematics and Computation}\
  }\textbf {\bibinfo {volume} {382}},\ \bibinfo {pages} {125331} (\bibinfo
  {year} {2020})}\BibitemShut {NoStop}%
\bibitem [{\citenamefont {Siu}\ \emph {et~al.}()\citenamefont {Siu},
  \citenamefont {Samorodnitsky}, \citenamefont {Yu},\ and\ \citenamefont
  {He}}]{Siu2025}%
  \BibitemOpen
  \bibfield  {author} {\bibinfo {author} {\bibfnamefont {C.}~\bibnamefont
  {Siu}}, \bibinfo {author} {\bibfnamefont {G.}~\bibnamefont {Samorodnitsky}},
  \bibinfo {author} {\bibfnamefont {C.~L.}\ \bibnamefont {Yu}},\ and\ \bibinfo
  {author} {\bibfnamefont {R.}~\bibnamefont {He}},\ }\bibfield  {title}
  {\bibinfo {title} {The asymptotics of the expected betti numbers of
  preferential attachment clique complexes},\ }\href
  {https://doi.org/10.1017/apr.2024.66} {\bibfield  {journal} {\bibinfo
  {journal} {Advances in Applied Probability}\ }\textbf {\bibinfo {volume}
  {57}},\ \bibinfo {pages} {940–968}}\BibitemShut {NoStop}%
\bibitem [{\citenamefont {Petri}\ \emph {et~al.}(2014)\citenamefont {Petri},
  \citenamefont {Expert}, \citenamefont {Turkheimer}, \citenamefont
  {Carhart-Harris}, \citenamefont {Nutt}, \citenamefont {Hellyer},\ and\
  \citenamefont {Vaccarino}}]{petri2014homological}%
  \BibitemOpen
  \bibfield  {author} {\bibinfo {author} {\bibfnamefont {G.}~\bibnamefont
  {Petri}}, \bibinfo {author} {\bibfnamefont {P.}~\bibnamefont {Expert}},
  \bibinfo {author} {\bibfnamefont {F.}~\bibnamefont {Turkheimer}}, \bibinfo
  {author} {\bibfnamefont {R.}~\bibnamefont {Carhart-Harris}}, \bibinfo
  {author} {\bibfnamefont {D.}~\bibnamefont {Nutt}}, \bibinfo {author}
  {\bibfnamefont {P.~J.}\ \bibnamefont {Hellyer}},\ and\ \bibinfo {author}
  {\bibfnamefont {F.}~\bibnamefont {Vaccarino}},\ }\bibfield  {title} {\bibinfo
  {title} {Homological scaffolds of brain functional networks},\ }\href
  {https://doi.org/10.1098/rsif.2014.0873} {\bibfield  {journal} {\bibinfo
  {journal} {Journal of The Royal Society Interface}\ }\textbf {\bibinfo
  {volume} {11}},\ \bibinfo {pages} {20140873} (\bibinfo {year}
  {2014})}\BibitemShut {NoStop}%
\bibitem [{\citenamefont {Courtney}\ and\ \citenamefont
  {Bianconi}(2016)}]{courtney2016generalized}%
  \BibitemOpen
  \bibfield  {author} {\bibinfo {author} {\bibfnamefont {O.~T.}\ \bibnamefont
  {Courtney}}\ and\ \bibinfo {author} {\bibfnamefont {G.}~\bibnamefont
  {Bianconi}},\ }\bibfield  {title} {\bibinfo {title} {Generalized network
  structures: The configuration model and the canonical ensemble of simplicial
  complexes},\ }\href {https://doi.org/10.1103/PhysRevE.93.062311} {\bibfield
  {journal} {\bibinfo  {journal} {Phys. Rev. E}\ }\textbf {\bibinfo {volume}
  {93}},\ \bibinfo {pages} {062311} (\bibinfo {year} {2016})}\BibitemShut
  {NoStop}%
\bibitem [{\citenamefont {Sizemore}\ \emph {et~al.}(2017)\citenamefont
  {Sizemore}, \citenamefont {Giusti},\ and\ \citenamefont
  {Bassett}}]{sizemore2017classification}%
  \BibitemOpen
  \bibfield  {author} {\bibinfo {author} {\bibfnamefont {A.}~\bibnamefont
  {Sizemore}}, \bibinfo {author} {\bibfnamefont {C.}~\bibnamefont {Giusti}},\
  and\ \bibinfo {author} {\bibfnamefont {D.~S.}\ \bibnamefont {Bassett}},\
  }\bibfield  {title} {\bibinfo {title} {Classification of weighted networks
  through mesoscale homological features},\ }\href
  {https://doi.org/10.1093/comnet/cnw013} {\bibfield  {journal} {\bibinfo
  {journal} {Journal of Complex Networks}\ }\textbf {\bibinfo {volume} {5}},\
  \bibinfo {pages} {245} (\bibinfo {year} {2017})}\BibitemShut {NoStop}%
\bibitem [{\citenamefont {Adami}\ \emph {et~al.}(2025)\citenamefont {Adami},
  \citenamefont {Masoomy},\ and\ \citenamefont {Najafi}}]{Adami_2025}%
  \BibitemOpen
  \bibfield  {author} {\bibinfo {author} {\bibfnamefont {V.}~\bibnamefont
  {Adami}}, \bibinfo {author} {\bibfnamefont {H.}~\bibnamefont {Masoomy}},\
  and\ \bibinfo {author} {\bibfnamefont {M.~N.}\ \bibnamefont {Najafi}},\
  }\bibfield  {title} {\bibinfo {title} {Topological data analysis of the
  visibility graphs of the btw avalanche-size sequences},\ }\href
  {https://doi.org/10.1088/1751-8121/add4d4} {\bibfield  {journal} {\bibinfo
  {journal} {Journal of Physics A: Mathematical and Theoretical}\ }\textbf
  {\bibinfo {volume} {58}},\ \bibinfo {pages} {205002} (\bibinfo {year}
  {2025})}\BibitemShut {NoStop}%
\bibitem [{\citenamefont {Bianconi}\ and\ \citenamefont
  {Rahmede}(2016)}]{bianconi2016network}%
  \BibitemOpen
  \bibfield  {author} {\bibinfo {author} {\bibfnamefont {G.}~\bibnamefont
  {Bianconi}}\ and\ \bibinfo {author} {\bibfnamefont {C.}~\bibnamefont
  {Rahmede}},\ }\bibfield  {title} {\bibinfo {title} {Network geometry with
  flavor: from complexity to quantum geometry},\ }\href
  {https://doi.org/10.1103/PhysRevE.93.032315} {\bibfield  {journal} {\bibinfo
  {journal} {Physical Review E}\ }\textbf {\bibinfo {volume} {93}},\ \bibinfo
  {pages} {032315} (\bibinfo {year} {2016})}\BibitemShut {NoStop}%
\bibitem [{\citenamefont {Bianconi}\ and\ \citenamefont
  {Rahmede}(2017)}]{bianconi2017emergent}%
  \BibitemOpen
  \bibfield  {author} {\bibinfo {author} {\bibfnamefont {G.}~\bibnamefont
  {Bianconi}}\ and\ \bibinfo {author} {\bibfnamefont {C.}~\bibnamefont
  {Rahmede}},\ }\bibfield  {title} {\bibinfo {title} {Emergent hyperbolic
  network geometry},\ }\href {https://doi.org/10.1038/srep41974} {\bibfield
  {journal} {\bibinfo  {journal} {Scientific reports}\ }\textbf {\bibinfo
  {volume} {7}},\ \bibinfo {pages} {1} (\bibinfo {year} {2017})}\BibitemShut
  {NoStop}%
\bibitem [{\citenamefont {Kovalenko}\ \emph {et~al.}(2021)\citenamefont
  {Kovalenko}, \citenamefont {Sendi{\~n}a-Nadal}, \citenamefont {Khalil},
  \citenamefont {Dainiak}, \citenamefont {Musatov}, \citenamefont
  {Raigorodskii}, \citenamefont {Alfaro-Bittner}, \citenamefont {Barzel},\ and\
  \citenamefont {Boccaletti}}]{kovalenko2021growing}%
  \BibitemOpen
  \bibfield  {author} {\bibinfo {author} {\bibfnamefont {K.}~\bibnamefont
  {Kovalenko}}, \bibinfo {author} {\bibfnamefont {I.}~\bibnamefont
  {Sendi{\~n}a-Nadal}}, \bibinfo {author} {\bibfnamefont {N.}~\bibnamefont
  {Khalil}}, \bibinfo {author} {\bibfnamefont {A.}~\bibnamefont {Dainiak}},
  \bibinfo {author} {\bibfnamefont {D.}~\bibnamefont {Musatov}}, \bibinfo
  {author} {\bibfnamefont {A.~M.}\ \bibnamefont {Raigorodskii}}, \bibinfo
  {author} {\bibfnamefont {K.}~\bibnamefont {Alfaro-Bittner}}, \bibinfo
  {author} {\bibfnamefont {B.}~\bibnamefont {Barzel}},\ and\ \bibinfo {author}
  {\bibfnamefont {S.}~\bibnamefont {Boccaletti}},\ }\bibfield  {title}
  {\bibinfo {title} {Growing scale-free simplices},\ }\href
  {https://doi.org/10.1038/s42005-021-00538-y} {\bibfield  {journal} {\bibinfo
  {journal} {Communications Physics}\ }\textbf {\bibinfo {volume} {4}},\
  \bibinfo {pages} {1} (\bibinfo {year} {2021})}\BibitemShut {NoStop}%
\bibitem [{\citenamefont {Kahle}(2009)}]{KAHLE20091658}%
  \BibitemOpen
  \bibfield  {author} {\bibinfo {author} {\bibfnamefont {M.}~\bibnamefont
  {Kahle}},\ }\bibfield  {title} {\bibinfo {title} {Topology of random clique
  complexes},\ }\href
  {https://doi.org/https://doi.org/10.1016/j.disc.2008.02.037} {\bibfield
  {journal} {\bibinfo  {journal} {Discrete Mathematics}\ }\textbf {\bibinfo
  {volume} {309}},\ \bibinfo {pages} {1658} (\bibinfo {year}
  {2009})}\BibitemShut {NoStop}%
\bibitem [{\citenamefont {Sizemore}\ \emph {et~al.}(2019)\citenamefont
  {Sizemore}, \citenamefont {Phillips-Cremins}, \citenamefont {Ghrist},\ and\
  \citenamefont {Bassett}}]{sizemore2019importance}%
  \BibitemOpen
  \bibfield  {author} {\bibinfo {author} {\bibfnamefont {A.~E.}\ \bibnamefont
  {Sizemore}}, \bibinfo {author} {\bibfnamefont {J.~E.}\ \bibnamefont
  {Phillips-Cremins}}, \bibinfo {author} {\bibfnamefont {R.}~\bibnamefont
  {Ghrist}},\ and\ \bibinfo {author} {\bibfnamefont {D.~S.}\ \bibnamefont
  {Bassett}},\ }\bibfield  {title} {\bibinfo {title} {The importance of the
  whole: Topological data analysis for the network neuroscientist},\ }\href
  {https://doi.org/10.1162/netn_a_00073} {\bibfield  {journal} {\bibinfo
  {journal} {Network Neuroscience}\ }\textbf {\bibinfo {volume} {3}},\ \bibinfo
  {pages} {656} (\bibinfo {year} {2019})}\BibitemShut {NoStop}%
\bibitem [{\citenamefont {Kahle}(2014)}]{kahle2014sharp}%
  \BibitemOpen
  \bibfield  {author} {\bibinfo {author} {\bibfnamefont {M.}~\bibnamefont
  {Kahle}},\ }\bibfield  {title} {\bibinfo {title} {Sharp vanishing thresholds
  for cohomology of random flag complexes},\ }\href
  {http://www.jstor.org/stable/24522785} {\bibfield  {journal} {\bibinfo
  {journal} {Annals of Mathematics}\ }\textbf {\bibinfo {volume} {179}},\
  \bibinfo {pages} {1085} (\bibinfo {year} {2014})}\BibitemShut {NoStop}%
\bibitem [{\citenamefont {Nakahara}(2018)}]{nakahara2018geometry}%
  \BibitemOpen
  \bibfield  {author} {\bibinfo {author} {\bibfnamefont {M.}~\bibnamefont
  {Nakahara}},\ }\href {https://doi.org/10.1201/9781315275826} {\emph {\bibinfo
  {title} {Geometry, topology and physics}}}\ (\bibinfo  {publisher} {CRC
  press},\ \bibinfo {year} {2018})\BibitemShut {NoStop}%
\bibitem [{\citenamefont {Dorogovtsev}\ and\ \citenamefont
  {Mendes}(2002)}]{Dorogovtsev01062002}%
  \BibitemOpen
  \bibfield  {author} {\bibinfo {author} {\bibfnamefont {S.~N.}\ \bibnamefont
  {Dorogovtsev}}\ and\ \bibinfo {author} {\bibfnamefont {J.~F.~F.}\
  \bibnamefont {Mendes}},\ }\bibfield  {title} {\bibinfo {title} {Evolution of
  networks},\ }\href {https://doi.org/10.1080/00018730110112519} {\bibfield
  {journal} {\bibinfo  {journal} {Advances in Physics}\ }\textbf {\bibinfo
  {volume} {51}},\ \bibinfo {pages} {1079} (\bibinfo {year}
  {2002})}\BibitemShut {NoStop}%
\bibitem [{\citenamefont {Carlsson}(2009)}]{carlsson2009topology}%
  \BibitemOpen
  \bibfield  {author} {\bibinfo {author} {\bibfnamefont {G.}~\bibnamefont
  {Carlsson}},\ }\bibfield  {title} {\bibinfo {title} {Topology and data},\
  }\href@noop {} {\bibfield  {journal} {\bibinfo  {journal} {Bulletin of the
  American Mathematical Society}\ }\textbf {\bibinfo {volume} {46}},\ \bibinfo
  {pages} {255} (\bibinfo {year} {2009})}\BibitemShut {NoStop}%
\bibitem [{\citenamefont {Adami}\ \emph {et~al.}(2026)\citenamefont {Adami},
  \citenamefont {Emdadi-Mahdimahalleh}, \citenamefont {Herrmann},\ and\
  \citenamefont {Najafi}}]{adami2026centrality}%
  \BibitemOpen
  \bibfield  {author} {\bibinfo {author} {\bibfnamefont {V.}~\bibnamefont
  {Adami}}, \bibinfo {author} {\bibfnamefont {S.}~\bibnamefont
  {Emdadi-Mahdimahalleh}}, \bibinfo {author} {\bibfnamefont {H.}~\bibnamefont
  {Herrmann}},\ and\ \bibinfo {author} {\bibfnamefont {M.}~\bibnamefont
  {Najafi}},\ }\bibfield  {title} {\bibinfo {title} {Centrality and
  universality in scale-free networks},\ }\href
  {https://doi.org/10.1103/tsmc-cnmz} {\bibfield  {journal} {\bibinfo
  {journal} {Physical Review E}\ }\textbf {\bibinfo {volume} {113}},\ \bibinfo
  {pages} {024307} (\bibinfo {year} {2026})}\BibitemShut {NoStop}%
\bibitem [{\citenamefont {V\'azquez}\ \emph {et~al.}(2002)\citenamefont
  {V\'azquez}, \citenamefont {Pastor-Satorras},\ and\ \citenamefont
  {Vespignani}}]{vazquez2001large}%
  \BibitemOpen
  \bibfield  {author} {\bibinfo {author} {\bibfnamefont {A.}~\bibnamefont
  {V\'azquez}}, \bibinfo {author} {\bibfnamefont {R.}~\bibnamefont
  {Pastor-Satorras}},\ and\ \bibinfo {author} {\bibfnamefont {A.}~\bibnamefont
  {Vespignani}},\ }\bibfield  {title} {\bibinfo {title} {Large-scale
  topological and dynamical properties of the internet},\ }\href
  {https://doi.org/10.1103/PhysRevE.65.066130} {\bibfield  {journal} {\bibinfo
  {journal} {Phys. Rev. E}\ }\textbf {\bibinfo {volume} {65}},\ \bibinfo
  {pages} {066130} (\bibinfo {year} {2002})}\BibitemShut {NoStop}%
\bibitem [{\citenamefont {Qi}\ and\ \citenamefont
  {Zhang}(2025)}]{qi2025iterative}%
  \BibitemOpen
  \bibfield  {author} {\bibinfo {author} {\bibfnamefont {L.}~\bibnamefont
  {Qi}}\ and\ \bibinfo {author} {\bibfnamefont {J.}~\bibnamefont {Zhang}},\
  }\href {https://arxiv.org/abs/2511.05869} {\bibinfo {title} {Iterative
  generation and generalized degree distribution of higher-order fractal
  scale-free networks}} (\bibinfo {year} {2025}),\ \Eprint
  {https://arxiv.org/abs/2511.05869} {arXiv:2511.05869 [math.CO]} \BibitemShut
  {NoStop}%
\bibitem [{\citenamefont {Xie}\ \emph {et~al.}(2022)\citenamefont {Xie},
  \citenamefont {Zhao},\ and\ \citenamefont {Yang}}]{xie2021higher}%
  \BibitemOpen
  \bibfield  {author} {\bibinfo {author} {\bibfnamefont {F.}~\bibnamefont
  {Xie}}, \bibinfo {author} {\bibfnamefont {Z.}~\bibnamefont {Zhao}},\ and\
  \bibinfo {author} {\bibfnamefont {P.}~\bibnamefont {Yang}},\ }\bibfield
  {title} {\bibinfo {title} {Higher order preference on the evolution of
  cooperation on barab{\'a}si--albert scale-free network},\ }in\ \href
  {https://doi.org/10.1007/978-3-030-89698-0_15} {\emph {\bibinfo {booktitle}
  {Advances in Natural Computation, Fuzzy Systems and Knowledge Discovery}}},\
  \bibinfo {editor} {edited by\ \bibinfo {editor} {\bibfnamefont
  {Q.}~\bibnamefont {Xie}}, \bibinfo {editor} {\bibfnamefont {L.}~\bibnamefont
  {Zhao}}, \bibinfo {editor} {\bibfnamefont {K.}~\bibnamefont {Li}}, \bibinfo
  {editor} {\bibfnamefont {A.}~\bibnamefont {Yadav}},\ and\ \bibinfo {editor}
  {\bibfnamefont {L.}~\bibnamefont {Wang}}}\ (\bibinfo  {publisher} {Springer
  International Publishing},\ \bibinfo {address} {Cham},\ \bibinfo {year}
  {2022})\ pp.\ \bibinfo {pages} {137--149}\BibitemShut {NoStop}%
\bibitem [{\citenamefont {Kannan}\ \emph {et~al.}(2019)\citenamefont {Kannan},
  \citenamefont {Saucan}, \citenamefont {Roy},\ and\ \citenamefont
  {Samal}}]{kannan2019persistent}%
  \BibitemOpen
  \bibfield  {author} {\bibinfo {author} {\bibfnamefont {H.}~\bibnamefont
  {Kannan}}, \bibinfo {author} {\bibfnamefont {E.}~\bibnamefont {Saucan}},
  \bibinfo {author} {\bibfnamefont {I.}~\bibnamefont {Roy}},\ and\ \bibinfo
  {author} {\bibfnamefont {A.}~\bibnamefont {Samal}},\ }\bibfield  {title}
  {\bibinfo {title} {Persistent homology of unweighted complex networks via
  discrete morse theory},\ }\href {https://doi.org/10.1038/s41598-019-50202-3}
  {\bibfield  {journal} {\bibinfo  {journal} {Scientific reports}\ }\textbf
  {\bibinfo {volume} {9}},\ \bibinfo {pages} {1} (\bibinfo {year}
  {2019})}\BibitemShut {NoStop}%
\bibitem [{\citenamefont {Roy}\ \emph {et~al.}(2020)\citenamefont {Roy},
  \citenamefont {Vijayaraghavan}, \citenamefont {Ramaia},\ and\ \citenamefont
  {Samal}}]{roy2020forman}%
  \BibitemOpen
  \bibfield  {author} {\bibinfo {author} {\bibfnamefont {I.}~\bibnamefont
  {Roy}}, \bibinfo {author} {\bibfnamefont {S.}~\bibnamefont {Vijayaraghavan}},
  \bibinfo {author} {\bibfnamefont {S.~J.}\ \bibnamefont {Ramaia}},\ and\
  \bibinfo {author} {\bibfnamefont {A.}~\bibnamefont {Samal}},\ }\bibfield
  {title} {\bibinfo {title} {Forman-ricci curvature and persistent homology of
  unweighted complex networks},\ }\href
  {https://doi.org/10.1016/j.chaos.2020.110260} {\bibfield  {journal} {\bibinfo
   {journal} {Chaos, Solitons \& Fractals}\ }\textbf {\bibinfo {volume}
  {140}},\ \bibinfo {pages} {110260} (\bibinfo {year} {2020})}\BibitemShut
  {NoStop}%
\bibitem [{\citenamefont {Gomez-Gardenes}\ and\ \citenamefont
  {Moreno}(2004)}]{gomez2004local}%
  \BibitemOpen
  \bibfield  {author} {\bibinfo {author} {\bibfnamefont {J.}~\bibnamefont
  {Gomez-Gardenes}}\ and\ \bibinfo {author} {\bibfnamefont {Y.}~\bibnamefont
  {Moreno}},\ }\bibfield  {title} {\bibinfo {title} {Local versus global
  knowledge in the barab{\'a}si-albert scale-free network model},\ }\href
  {https://doi.org/10.1103/PhysRevE.69.037103} {\bibfield  {journal} {\bibinfo
  {journal} {Physical Review E}\ }\textbf {\bibinfo {volume} {69}},\ \bibinfo
  {pages} {037103} (\bibinfo {year} {2004})}\BibitemShut {NoStop}%
\bibitem [{\citenamefont {Lee}\ \emph {et~al.}(2006)\citenamefont {Lee},
  \citenamefont {Kim},\ and\ \citenamefont {Jeong}}]{lee2006statistical}%
  \BibitemOpen
  \bibfield  {author} {\bibinfo {author} {\bibfnamefont {S.~H.}\ \bibnamefont
  {Lee}}, \bibinfo {author} {\bibfnamefont {P.-J.}\ \bibnamefont {Kim}},\ and\
  \bibinfo {author} {\bibfnamefont {H.}~\bibnamefont {Jeong}},\ }\bibfield
  {title} {\bibinfo {title} {Statistical properties of sampled networks},\
  }\href {https://doi.org/10.1103/PhysRevE.73.016102} {\bibfield  {journal}
  {\bibinfo  {journal} {Physical review E}\ }\textbf {\bibinfo {volume} {73}},\
  \bibinfo {pages} {016102} (\bibinfo {year} {2006})}\BibitemShut {NoStop}%
\bibitem [{\citenamefont {Holme}\ and\ \citenamefont
  {Kim}(2002)}]{holme2002growing}%
  \BibitemOpen
  \bibfield  {author} {\bibinfo {author} {\bibfnamefont {P.}~\bibnamefont
  {Holme}}\ and\ \bibinfo {author} {\bibfnamefont {B.~J.}\ \bibnamefont
  {Kim}},\ }\bibfield  {title} {\bibinfo {title} {Growing scale-free networks
  with tunable clustering},\ }\href
  {https://doi.org/10.1103/PhysRevE.65.026107} {\bibfield  {journal} {\bibinfo
  {journal} {Physical review E}\ }\textbf {\bibinfo {volume} {65}},\ \bibinfo
  {pages} {026107} (\bibinfo {year} {2002})}\BibitemShut {NoStop}%
\bibitem [{\citenamefont {Barab{\'a}si}\ \emph {et~al.}(1999)\citenamefont
  {Barab{\'a}si}, \citenamefont {Albert},\ and\ \citenamefont
  {Jeong}}]{barabasi1999mean}%
  \BibitemOpen
  \bibfield  {author} {\bibinfo {author} {\bibfnamefont {A.-L.}\ \bibnamefont
  {Barab{\'a}si}}, \bibinfo {author} {\bibfnamefont {R.}~\bibnamefont
  {Albert}},\ and\ \bibinfo {author} {\bibfnamefont {H.}~\bibnamefont
  {Jeong}},\ }\bibfield  {title} {\bibinfo {title} {Mean-field theory for
  scale-free random networks},\ }\href
  {https://doi.org/10.1016/S0378-4371(99)00291-5} {\bibfield  {journal}
  {\bibinfo  {journal} {Physica A: Statistical Mechanics and its Applications}\
  }\textbf {\bibinfo {volume} {272}},\ \bibinfo {pages} {173} (\bibinfo {year}
  {1999})}\BibitemShut {NoStop}%
\bibitem [{\citenamefont {Klemm}\ and\ \citenamefont
  {Eguiluz}(2002)}]{klemm2002growing}%
  \BibitemOpen
  \bibfield  {author} {\bibinfo {author} {\bibfnamefont {K.}~\bibnamefont
  {Klemm}}\ and\ \bibinfo {author} {\bibfnamefont {V.~M.}\ \bibnamefont
  {Eguiluz}},\ }\bibfield  {title} {\bibinfo {title} {Growing scale-free
  networks with small-world behavior},\ }\href
  {https://doi.org/10.1103/PhysRevE.65.057102} {\bibfield  {journal} {\bibinfo
  {journal} {Physical Review E}\ }\textbf {\bibinfo {volume} {65}},\ \bibinfo
  {pages} {057102} (\bibinfo {year} {2002})}\BibitemShut {NoStop}%
\bibitem [{\citenamefont {Knopp}(1990)}]{knopp1990theory}%
  \BibitemOpen
  \bibfield  {author} {\bibinfo {author} {\bibfnamefont {K.}~\bibnamefont
  {Knopp}},\ }\href {https://books.google.com/books?id=drJNXrAa7poC} {\emph
  {\bibinfo {title} {Theory and Application of Infinite Series}}},\ Dover Books
  on Mathematics\ (\bibinfo  {publisher} {Dover Publications},\ \bibinfo {year}
  {1990})\BibitemShut {NoStop}%
\bibitem [{\citenamefont {Morozov}()}]{dmitriydionysus}%
  \BibitemOpen
  \bibfield  {author} {\bibinfo {author} {\bibfnamefont {D.}~\bibnamefont
  {Morozov}},\ }\bibfield  {title} {\bibinfo {title} {Dionysus 2: the second
  incarnation of the library for computing persistent homology},\ }\href
  {https://mrzv.org/software/dionysus2/} {\bibinfo  {journal} {mrzv.org}\
  }\BibitemShut {NoStop}%
\end{thebibliography}%
	
\end{document}